\documentclass[aps,preprint,superscriptaddress,nofootinbib,preprintnumbers,prd,eqsecnum]{revtex4-2}

\usepackage{bm,amsmath,amsfonts,amssymb,slashed,graphicx,xspace,placeins,rotating}
\usepackage{xcolor,colortbl}
\usepackage{multirow,array,makecell}
\usepackage[inline]{enumitem}
\usepackage{xparse}
\allowdisplaybreaks
\DeclareMathAlphabet{\mathpzc}{OT1}{pzc}{m}{it}

\definecolor{nicered}{rgb}{0.7,0.1,0.1}
\definecolor{nicegreen}{rgb}{0.1,0.5,0.1}
\definecolor{niceblue}{rgb}{0.1,0.1,0.5}
\usepackage[bookmarks=false]{hyperref}
\hypersetup{colorlinks,citecolor= nicegreen,linkcolor= niceblue,urlcolor= niceblue}
\usepackage{breakurl}

\newcommand{\nn}{\nonumber}
\newcommand{\GeV}{\text{GeV}}

\def\d{{\rm d}}

\newcommand{\g}{\gamma}

\newcommand{\ab}{\alpha\beta}
\newcommand{\ceq}{\stackrel{\to}{=}}

\newcommand{\mubc}{\mu_{bc}}

\newcommand{\genbar}[1]{\,\overline{\!#1}{}}
\newcommand{\Overrightarrow}[1]{{%
	#1
}}
\newcommand{\Overleftarrow}[1]{{%
	#1
}}

\newcommand{\bbar}{\bar{b}}
\newcommand{\cbar}{\bar{c}}
\newcommand{\Lb}{\Lambda_b}
\newcommand{\Lc}{\Lambda_c}
\newcommand{\Lq}{\Lambda_Q}
\newcommand{\Bbar}{\,\overline{\!B}}

\newcommand{\cbvp}{\cbar^{v'}_+}

\newcommand{\bv}{b^v_+}
\newcommand{\bbv}{\bbar^v_+}

\newcommand{\Dslash}{\slashed{D}}
\newcommand{\vslash}{\slashed{v}}
\newcommand{\ccdot}{\!\cdot\!}
\newcommand{\vcD}{v \ccdot D}
\newcommand{\Qbar}{\genbar{Q}}

\newcommand{\mJ}{\mathcal{J}}
\newcommand{\mJbar}{\genbar{\mJ}}
\newcommand{\Hc}{H_c}
\newcommand{\Hb}{H_b}

\newcommand{\Hbar}{\genbar{H}}

\newcommand{\Us}{\mathcal{U}}
\newcommand{\Ub}{\genbar{\mathcal{U}}}

\newcommand{\Cs}{C_{S}}
\newcommand{\Cps}{C_{P}}
\newcommand{\Cv}[1]{C_{V#1}}
\newcommand{\Ca}[1]{C_{A#1}}
\newcommand{\Ct}[1]{C_{T#1}}

\newcommand{\hK}[2]{\hat{K}_{#1}^{(#2)}}
\newcommand{\hM}[1]{\hat{M}_{#1}}

\newcommand{\lqcd}{\Lambda_{\text{QCD}}}
\newcommand{\LamB}{\bar{\Lambda}_\Lambda}
\newcommand{\lam}[1]{\lambda^\Lambda_{#1}}
\NewDocumentCommand{\lamL}{ O{} O{} }{\frac{#1\lam1}{#2\LamB^2}}
\newcommand{\rh}[1]{\rho^\Lambda_{#1}}
\newcommand{\aS}{\alpha_s}
\newcommand{\haS}{{\hat{\alpha}_s}}
\newcommand{\ec}{\varepsilon_c}
\newcommand{\eb}{\varepsilon_b}
\newcommand{\eQ}{\varepsilon_Q}

\newcommand{\quot}{\natural}
\newcommand{\hvph}{\hat\varphi_1}
\newcommand{\hvphq}{\hat\varphi^\quot_1}
\newcommand{\hvphp}{\hat\varphi'_1}
\newcommand{\hvphpp}{\hat\varphi''_1}

\newcommand{\mbS}{m_b^{1S}}
\newcommand{\mcS}{m_c^{1S}}
\newcommand{\dmbc}{\delta m_{bc}}

\newcommand{\amp}[3]{\frac{\langle #1(p') |\, #2\, | #3(p) \rangle}{\sqrt{m_{\let\overline\relax#3} m_{#1}}}}

\newcommand{\nax}{\text{---}}

\newcommand{\Tr}{\text{Tr}}
\newcommand{\sqrtb}[1]{\sqrt{\smash[b]{#1}}\vphantom{b}}

\newcommand{\basescen}[1][]{baseline#1\xspace}
\newcommand{\blrsscen}[1][]{BLRS#1\xspace}
\newcommand{\lqcdscen}{LQCD only}
\newcommand{\LOcubic}{+LO3}

\makeatletter
\g@addto@macro\bfseries{\boldmath}
\makeatother

\makeatletter
\let\tmp@footnote\footnote
\renewcommand{\footnote}[1]{\tmp@footnote{\linespread{0.9}\selectfont{}#1}}
\makeatother

\makeatletter
\let\temp@caption\caption
\renewcommand{\caption}[2][]{\temp@caption[#1]{\linespread{1.2}\selectfont{}#2}}
\makeatother

\allowdisplaybreaks
\begin{document}

\preprint{CALT-TH-2023-049}

\title{Appraising constrained second-order power corrections in HQET with $\Lambda_b \to \Lambda_c l \nu$}

\author{Florian U.\ Bernlochner}
\affiliation{Physikalisches Institut der Rheinischen Friedrich-Wilhelms-Universit\"at Bonn, 53115 Bonn, Germany}

\author{Michele Papucci}
\affiliation{Walter Burke Institute for Theoretical Physics, California Institute of Technology, Pasadena, CA 91125, USA}

\author{Dean J.\ Robinson}
\affiliation{Ernest Orlando Lawrence Berkeley National Laboratory, 
University of California, Berkeley, CA 94720, USA}
\affiliation{Berkeley Center for Theoretical Physics, 
Department of Physics,
University of California, Berkeley, CA 94720, USA}

\begin{abstract}
We derive the $\Lambda_b \to \Lambda_c$ form factors for the Standard Model and beyond at second order in Heavy Quark Effective Theory (HQET), 
applying the recently-proposed Residual Chiral Expansion (RCE) to reduce the set of unknown subsubleading hadronic functions to a single, highly-constrained function,
that is fully determined by hadron mass parameters at zero recoil.
We fit a form factor parametrization based on these results to all available Lattice QCD (LQCD) predictions and experimental data.
We find that the constrained and predictive structure of the form factors under the RCE is in excellent agreement with LQCD predictions and experimental data,
as well as prior HQET-based fits.
\end{abstract}

\maketitle
\newpage

\twocolumngrid
{
\fontsize{10}{8}\selectfont 
\columnsep20pt
	\tableofcontents
}
\onecolumngrid

\section{Introduction}

Semileptonic $\Lb \to \Lc l \nu$ decays provide a sensitive laboratory for exploring the behavior of second-order power corrections in Heavy Quark Effective Theory (HQET).
Unlike in the $B \to D^{(*)}$ system, in $\Lb \to \Lc$ transitions no new subleading Isgur-Wise functions---the 
form factors that describe HQET matrix elements---enter 
at first order in the heavy quark (HQ) power expansion~\cite{Georgi:1990ei, Falk:1992ws}.
Thus the first-order HQ power expansion for $\Lb \to \Lc$ matrix elements is highly constrained, 
and the breaking of the heavy quark symmetry (HQS) at second order 
can be explored without ``contamination'' from unknown first-order hadronic functions.
Refs.~\cite{Bernlochner:2018kxh,Bernlochner:2018bfn} exploited this property 
to probe the size of $\mathcal{O}(1/m_c^2)$ contributions
and obtain precise predictions for the lepton flavor universality violation (LFUV) ratio
\begin{equation}
	R(\Lc) = \Gamma[\Lb \to \Lc \tau \nu]/\Gamma[\Lb \to \Lc \ell \nu]\,,\qquad \ell = \mu\,,e\,,
\end{equation}
using a combined fit to LQCD calculations of the $\Lb \to \Lc$ form factors~\cite{Detmold:2015aaa} 
and LHCb measurements of the $\Lb \to \Lc \mu \nu$ differential shapes~\cite{Aaij:2017svr}.
The ``BLRS'' form factor parametrization~\cite{Bernlochner:2018kxh} developed therein considered HQET power corrections up to and including $\mathcal{O}(1/m_c^2)$,
using the long-known results of Ref.~\cite{Falk:1992ws}.
The combined fit demonstrated current data are already precise enough to allow the two-parameter system at $\mathcal{O}(1/m_c^2)$ in the BLRS parametrization 
to be recovered at more than $2.5\sigma$ from zero~\cite{Bernlochner:2018kxh}.

The role of second-order power corrections in HQET-based descriptions of exclusive $b \to c l \nu$ semileptonic decays is of particularly high importance.
At the percent-level precision (or better) anticipated in upcoming measurements,
second-order power corrections in the HQ expansion, which naively enter at the several percent level, 
may introduce sizeable sources of theoretical uncertainty in the description of in the $B \to D^{(*)} l \nu$ decays.
This systematically limits (future) precision measurements of the CKM matrix element $|V_{cb}|$
as well as precision predictions and measurements of the LFUV ratios, 
$R(D)$ and $R(D^*)$ (see e.g. the introductory discussion in Ref.~\cite{Bernlochner:2022ywh}).

In Ref.~\cite{Bernlochner:2022ywh} we recently postulated a
supplemental power counting within HQET---the residual chiral expansion (RCE)---that 
reduces the typically large and overcomplete number of subsubleading Isgur-Wise functions 
entering at second-order to a smaller, highly-constrained set.
The RCE is in effect a truncation scheme based on a conjecture that matrix elements 
containing more insertions of the transverse residual momentum operator acting on the HQ mass-subtracted states 
are suppressed compared to those matrix elements containing fewer such insertions.
This coincides, beyond a certain number of insertions, with those matrix elements involving a larger number of operator product Lagrangian insertions being suppressed compared to those involving fewer.
Truncating at second order in the RCE, at zero recoil the second-order power corrections in HQET for $\Bbar \to D^{(*)}$ transitions 
become constrained by hadron mass parameters, leading to zero recoil predictions that are in good agreement with Lattice QCD (LQCD) predictions.
Ref.~\cite{Bernlochner:2022ywh} further showed that RCE-based parametrizations of these form factors achieved excellent fits to LQCD calculations and Belle data.

In general the HQ power expansion for $\Lb \to \Lc$ matrix elements is simpler than in the $\Bbar \to D^{(*)}$ system
because the HQET describing these decays involve light degrees of freedom in the simplest spin-parity $s_\ell^{\pi_\ell} = 0^+$ state,
rather than $s_\ell^{\pi_\ell} =1/2^+$.
Noting the abovementioned (and related) absence of first-order Isgur-Wise functions in $\Lb \to \Lc$ matrix elements,
the $\Lb \to \Lc$ system therefore provides a particularly clean laboratory to explore the behavior of the RCE,
and to test whether RCE-based parametrizations and predictions are compatible with current LQCD predictions and available experimental data,
as well as prior BLRS precision fits. 
(Some exploration of RCE-based parametrizations for the more complicated $\Lb \to \Lc^*$ system has also been conducted, with promising results~\cite{DiRisi:2023npw}.)

In this work we show that at second order in the RCE,
the second-order power corrections are described entirely by a single subsubleading Isgur-Wise function, 
that is constrained at zero recoil by hadron mass parameters.
This holds not only at $\mathcal{O}(1/m_{c}^2)$ but at $\mathcal{O}(1/m_{c,b}^2)$ in the HQ power expansion.\footnote{Here 
and throughout this work we use $\mathcal{O}(1/m_{c,b}^2)$ to denote terms of order $\sim 1/m_c^2$, $1/m_cm_b$, and $1/m_b^2$.}
By comparison, six (three) subsubleading Isgur-Wise functions enter the full HQET description at second order ($\mathcal{O}(1/m_c^2)$),
describing six (two) HQS functions. 
A summary is provided in Table~\ref{tab:HQETFFs}; see also Sec.~\ref{sec:summff} for an explanation of this counting.

\begin{table}[t!]
\renewcommand*{\arraystretch}{0.9}
\newcolumntype{C}{ >{\centering\arraybackslash } m{2.5cm} <{}}
\begin{tabular}{ccCCC}
	\hline\hline
	\multirow{2}{*}{HQET order} & \multirow{2}{*}{\makecell[c]{Fixed-order \\[-5pt] HQS functions}} & \multicolumn{2}{c}{Isgur-Wise functions} \\
	&& All& RCE\\
	\hline
	$1/m_{c,b}^0$ 	& 1 & 1 & 1\\
	$1/m_{c,b}^1$	& 2 & 0 & 0 \\
	$1/m_c^2$  & 2 & 3  & 1\\
	$1/m_{c,b}^2$ & 6& 6& 1\\
 	\hline\hline
\end{tabular}
\caption{Number of $\Lb \to \Lc$ form factors, fixed-order HQS functions, and Isgur-Wise functions entering at each fixed order in HQET.}
\label{tab:HQETFFs}
\end{table}

We show that the resulting zero-recoil RCE-based predictions for the $f_1$ and $g_1$ SM form factors at $\mathcal{O}(\aS^2, \aS/m_{c,b}, 1/m_{c,b}^2)$ are in good agreement with LQCD calculations.
Further, we show that RCE-based form factor parametrizations at $\mathcal{O}(\aS^2, \aS/m_{c,b}, 1/m_{c,b}^2)$ yield fits 
in good agreement with LQCD predictions and LHCb data as well as prior BLRS fit results, 
despite the parametrization being more constrained compared to the standard HQET description.
The predictions for $R(\Lc)$ derived from these RCE-based fits are similarly compatible with BLRS results. 
In particular, for our ``\basescen'' fit scenario we recover
\begin{equation}
	R(\Lc) =  0.3251(37)\,,
\end{equation}
in good agreement with the BLRS result $R(\Lc) = 0.3233(37)$.\footnote{An adjustment considering the impact of bin correlations on the LHCb data normalization, 
along with a minor discrepancy in the $\aS^2$ terms in the $1S$ mass scheme used for the leading renormalon cancellation, 
introduces slight modifications to the BLRS fit. Consequently, the resulting BLRS prediction for $R(\Lambda_c)$ differs marginally 
from the previously reported value of $0.3237(36)$~\cite{Bernlochner:2018kxh}. See Sec.~\ref{sec:data}\label{ft:RLc} for more details.}
The paper is organized as follows: 
In Sec.~\ref{sec:prelim} we specialize the RCE to the $\Lb \to \Lc$ system.
In Sec.~\ref{sec:RCEformfact} and in Appendices~\ref{app:ffderivation} and~\ref{app:radcrr}
we compute the form factors at second order in the RCE and $\mathcal{O}(\aS/m_{c,b}, 1/m_{c,b}^2)$ in HQET,
both in the Standard Model (SM) and beyond.
Section~\ref{sec:RCEformfact} further presents the parameterizations of the leading and subsubleading Isgur-Wise functions, 
the application of $1S$ short distance mass scheme,
and the SM zero-recoil predictions at second order in the RCE compared to LQCD calculations. 
Section~\ref{sec:fits} presents our fit scenarios and  fit results to LQCD and LHCb data.
Section~\ref{sec:summ} concludes.

\section{RCE Preliminaries}\label{sec:prelim}
The following specializes the general discussion in Ref.~\cite{Bernlochner:2022ywh} for $b$- to~$c$-hadron transitions to the case of $\Lb \to \Lc$ transitions,
for which the matrix elements are described by matching onto the $s_\ell^{\pi_\ell} = 0^+$ HQET.
These transitions are mediated by the current $J_\Gamma = \cbar \,\Gamma\, b$, where $\Gamma$ denotes any Dirac operator.
A full operator basis entering the QCD matrix elements $\langle \Lc | \cbar \,\Gamma\, b | \Lb \rangle$ is chosen to be
\begin{equation}
	\label{eqn:Gdef}
	J_S = \bar c\, b\,, \quad  J_P = \bar c\, \g^5\, b\,, \quad
	J_V = \bar c\,\g^\mu\, b\,, \quad  J_A = \bar c\, \g^\mu\g^5\, b\,, \quad
	J_T = \bar c\, \sigma^{\mu\nu}\, b\,,
\end{equation}
where $\sigma^{\mu\nu} \equiv (i/2)[\g^\mu,\g^\nu]$.  
The pseudotensor contribution is determined by the identity $\sigma^{\mu\nu} \g^5 \equiv +(i/2)\epsilon^{\mu \nu \rho \sigma}
\sigma_{\rho \sigma}$, which implies $\text{Tr}[\g^\mu\g^\nu\g^\rho\g^\sigma\g^5] = -4i \epsilon^{\mu\nu\rho\sigma}$.
(An opposite sign convention is usually chosen for $\Bbar \to D^{(*)}$ transitions.)

Using the notation and results of Ref.~\cite{Bernlochner:2022ywh}, 
at second order in the HQ expansion
and keeping terms only to second order in the RCE---i.e. 
at ``$\mathcal{O}(\theta^2)$''---the 
matching between the $\Lb \to \Lc$ QCD and HQET matrix elements simplifies to
\begin{align}
	& \frac{\langle \Lc | \cbar \,\Gamma\, b | \Lb \rangle}{\sqrt{m_{\Lc} m_{\Lb}}}  
	 \simeq \big\langle \Lc^{v'} \big| \cbvp \, \Gamma \, \bv \big| \Lb^v \big\rangle \label{eqn:QCDmatchRC}\\*
	& \quad + \frac{1}{2m_c} \big\langle \Lc^{v'} \big|  \big(\cbvp\mJbar^{\prime}_1\Pi_-' + \mathcal{L}'_1 \circ \cbvp \big) \Gamma \, \bv \big| \Lb^v \big\rangle
	 + \frac{1}{2m_b} \big\langle \Lc^{v'} \big| \cbvp  \, \Gamma  \big(\Pi_-\mJ_1\bv + \bv \circ \mathcal{L}_1 \big)  \big| \Lb^v \big\rangle \nn \\
	& \quad + \frac{1}{4m_c^2} \big\langle \Lc^{v'} \big| \big(\cbvp\mJbar^{\prime}_2\Pi_-' + \mathcal{L}'_2 \circ \cbvp \big) \Gamma \, \bv \big| \Lb^v \big\rangle 
	  + \frac{1}{4m_b^2} \big\langle \Lc^{v'} \big| \cbvp \, \Gamma \big(\Pi_-\mJ_2\bv  + \bv \circ \mathcal{L}_2 \big) \big| \Lb^v \big\rangle \nn \\*
	& \quad + \frac{1}{4m_c m_b} \big\langle \Lc^{v'} \big| \cbvp \mJbar^{\prime}_1\Pi_-' 
		\Gamma \Pi_- \mJ_1 \bv \big| \Lb^v \big\rangle\,. \nn
\end{align}
Here $\Pi_\pm = (1 \pm \vslash)/2$, and $|\Lq^v\rangle$ denote the eigenstates of the leading order HQET Lagrangian,
with HQ velocity $v$.
These states are normalized such that 
\begin{equation}
	\label{eqn:HQETnorm}
	\langle \Lq^{v'}(k') | \Lq^v(k) \rangle = 2v^0 \delta_{v v'}(2\pi)^3 \delta^3(k-k')\,.
\end{equation}
Note this normalization choice differs from that in Ref.~\cite{Falk:1992ws}, which normalized the HQET states with respect to an HQ mass scale, $m_Q + \LamB$
(see Eq.~\eqref{eqn:HmassHQE} below).
In Eq.~\eqref{eqn:QCDmatchRC}, the Lagrangian corrections
\begin{subequations}
\begin{align}
	\mathcal{L}_1 & = -\Qbar^v_+ \Dslash_{\perp} \Dslash_{\perp} Q^v_+  =  -\Qbar^v_+ \bigg[D^2 + a_Q(\mu)\frac{g}{2} \sigma_{\ab} G^{\ab} \bigg]Q^v_+ \,, \label{eqn:mL1}\\*
	\mathcal{L}_2 & =  \Qbar^v_+  [\Dslash_{\perp} i\vcD \Dslash_{\perp} ] Q^v_+  =  g\Qbar^v_+ \bigg[ v_\beta  D_\alpha G^{\ab}  - i v_\alpha \sigma_{\beta\gamma} D^\gamma G^{\ab} \bigg]Q^v_+ \,, \label{eqn:mL2}
\end{align}
\end{subequations}
and the current corrections $\mJ_1 = i \Dslash$, and $\mJ_2 = -\Dslash \Dslash$. 
The conjugate forms $\mJbar_n \equiv \gamma^0 \overleftarrow{\mJ}_n^\dagger \gamma^0$, where the the arrow indicates action of the derivatives to the left.
The $\circ$ operator in Eq.~\eqref{eqn:QCDmatchRC} denotes a (time-ordered) operator product, e.g.  $\mathcal{L}'_1 \circ \cbvp(z)  = i\int d^4 x \,  \mathcal{L}'_1(x) \cbvp(z)$.

In the $s_\ell^P = 0^+$ HQET, the matrix elements generated by the chromomagnetic terms in $\mathcal{L}_1$ and $\mathcal{L}_2$ vanish (see Appendix~\ref{app:ffderivation}).
Related to this, matching of the QCD correlator involving the HQET Hamiltonian $\langle \Lq \big| \Qbar^v_+ i \vcD Q^v_+ \big| \Lq \rangle$ onto HQET
yields the hadron mass expansion
\begin{equation}
	\label{eqn:HmassHQE}
	m_{\Lq} = m_Q + \LamB + \frac{\Delta m_2}{2m_Q} + \ldots\,, \qquad \Delta m_2 = - \lam1\,.
\end{equation}
The hadron mass parameter $\LamB$ corresponds to the kinetic energy of the brown muck in the HQ symmetry limit, 
while $\lam1  = -\langle\Lq^v | \Qbar_+^v D^2 Q^v_+ | \Lq^v \rangle/2$ is generated by the kinetic energy term in $\mathcal{L}_1$.

Contact terms generated by derivatives acting on HQET matrix elements give rise to Schwinger-Dyson relations
(sometimes called modified Ward identities).
These relate matrix elements entering at different orders in the HQ expansion~\eqref{eqn:QCDmatchRC}, 
reducing the total number of unknown hadronic functions entering at a particular order.
In this work, we always choose the HQ velocity to be that of the hadron containing it, i.e., $p = m_H v$.
At leading order, one then has the (familiar) Schwinger-Dyson relation
\begin{equation}
	\label{eqn:SDrelNLO}
	\big\langle \Lc^{v'} \big|  \cbvp(z) i\overleftarrow{D}^z_\mu \, \Gamma \, \bv(z)  \big| \Lb^v \big\rangle 
	+ \big\langle \Lc^{v'} \big|  \cbvp(z) \, \Gamma i\overrightarrow{D}^z_\mu  \bv(z) \big| \Lb^v \big\rangle
	\ceq \LamB(v-v')_\mu \big\langle \Lc^{v'} \big| \cbvp(z) \Gamma \bv(z) \big| \Lb^v \big\rangle\,,
\end{equation}
where the operator `$\ceq$' indicates a relation that arises under composition of the hadron current in question with an external operator, 
using integration by parts and enforcing overall momentum conservation.
At next-to-leading order, and keeping only terms to $\mathcal{O}(\theta^2)$, 
the Schwinger-Dyson relations become
\begin{align}
	\label{eqn:SDrelNNLO}
	\big\langle \Lc^{v'} \big|  \cbvp(z) \Overleftarrow{\mJbar}^{\prime}_2 \, \Pi'_+ \Gamma \, \bv(z)  \big| \Lb^v \big\rangle 
	& \ceq \lam1 \big\langle \Lc^{v'} \big| \cbvp(z) \Gamma \bv(z) \big| \Lb^v \big\rangle\,\nn\\*
	\big\langle \Lc^{v'} \big|  \cbvp(z) \Gamma \, \Pi_+ \Overrightarrow{\mJ}_2 \, \bv(z)  \big| \Lb^v \big\rangle 
	& \ceq \lam1 \big\langle \Lc^{v'} \big| \cbvp(z) \Gamma \bv(z) \big| \Lb^v \big\rangle\,.
\end{align}

\section{$\Lb \to \Lc$ form factors}
\label{sec:RCEformfact}

\subsection{HQET matrix elements}
The $\Lq$ ground-state baryons are formed by the tensor product of a spin-$1/2$ heavy quark 
with brown muck in the spin-parity $s_\ell^{\pi_\ell} = 0^+$ definite state, and thus belong to a HQ singlet.
The particle representation of this singlet is a Dirac spinor $\Us^v$, obeying the Dirac equation of motion $\vslash\,\Us^v = \Us^v$.

The matching of the QCD matrix elements onto HQET in Eq.~\eqref{eqn:QCDmatchRC} then becomes 
\begin{align}
	\frac{\langle \Lc(p') | \cbar \,\Gamma\, b | \Lb(p) \rangle}{\sqrt{m_{\Lc} m_{\Lb}}}  
	& = \zeta(w) \bigg\{ \Ub^{v'} \,\Gamma\, \Us^v \nn \\*
	& + \sum_{n=1}^2 \ec^n\, \Ub^{(n)}(v',v) \,\Gamma\, \Us^v  + \sum_{n=1}^2 \eb^n\,  \Ub^{v'} \,\Gamma\, \Us^{(n)}(v,v')  \nn\\*
	& + \ec\eb \Tr\big[   \Gamma\, \mathcal{W}^{(1,1)}(v,v')\big]\bigg\}\,, \label{eqn:LbLcmatch}
\end{align}
choosing HQ velocities $v = p/m_{\Lb}$, $v' = p'/m_{\Lc}$, and defining
\begin{align}
	\Us^{(n)}(v,v')  & = \hK1{n}\, \Us^v + \hK2{n}\, \Pi_- \vslash' \Us^v\,, \\
	\mathcal{W}^{(1,1)}(v,v') & = \hM1 \, \Us^v \Ub^{v'} 
		+ \hM2 \, \Pi_- \vslash' \Us^v \Ub^{v'} + \hM2'\,  \Us^v \Ub^{v'} \vslash \Pi_- \nn \\
		& \qquad + \hM3 \, \Pi_- \vslash' \Us^v \Ub^{v'} \vslash \Pi_- + \hM4 \, \Pi_- \gamma_\alpha \Us^v \Ub^{v'} \gamma^\alpha \Pi_- \,,
\end{align}
in which $\hK1{n}$ and $\hM{i}$ are ``HQS functions'' of the recoil parameter
\begin{equation}
	w = v \ccdot v' = \frac{m_{\Lb}^2 + m_{\Lc}^2 - q^2}{2 m_{\Lb} m_{\Lc}}\,, \qquad q^2 = (p -p')^2\,.
\end{equation}
The heavy quark expansion parameters $\varepsilon_{c,b} = \LamB/(2m_{c,b})$,
such that all $\hK{i}{n}$, $\hM{i}$, and Isgur-Wise functions are dimensionless.
Note also $\hM2 = \hM2'$ up to $\mathcal{O}(\aS \times 1/m_cm_b)$, i.e. third-order, perturbative corrections, which can be included separately.

In Eq.~\eqref{eqn:LbLcmatch} we have factored out the leading Isgur-Wise function, $\zeta(w)$, from all terms.
We use the notation that hatted functions of $w$ are normalized to $\zeta(w)$, i.e. $\hat{W}(w) \equiv W(w)/\zeta(w)$ for any Isgur-Wise function or form factor, $W(w)$.
In the equal mass, zero-recoil limit the normalization of the QCD matrix element for the (conserved) vector current requires that
\begin{equation}
	\label{eqn:massnormcon}
	\langle \Lq | \Qbar \g^\mu Q | \Lq \rangle = 2 m_{\Lq} v^\mu\,.
\end{equation}
When matched onto HQET at leading order, this implies that $\zeta(1) = 1$. 
Further zero-recoil constraints are discussed below in Sec.~\ref{sec:summff}.

Terms generated by current corrections in Eq.~\eqref{eqn:LbLcmatch} are always associated with the presence of a $\Pi_-^{(\prime)}$ projector acting on the (anti)spinor
while Lagrangian corrections are always associated with a $\Pi_+^{(\prime)}$: see e.g.~Eq.~\eqref{eqn:QCDmatchRC}.
Thus, we expect $\hK1{n}$ to contain Lagrangian corrections only, and $\hK2{n}$ to contain only current corrections.
Further, since in the RCE at $\mathcal{O}(1/m_cm_b,\theta^2)$ only the matrix element 
generated by the product current correction $\sim \mJbar_1'  \mJ_1$ is present, 
then we expect that $\hM1$ and $\hM2$ will vanish, but $\hM3$ and $\hM4$ will remain non-zero.

\subsection{Form factor matching}
Following the standard HQET-style basis for the $\Lb \to \Lc$ form factors,
the SM matrix elements can be represented as~\cite{Isgur:1990pm, Falk:1992ws, Manohar:2000dt}
\begin{align}
\label{eqn:HQETffdefSM}
\langle \Lc(p',s')| \bar c\gamma_\nu b |\Lb(p,s)\rangle
  &= \bar u(p',s') \big[ f_1 \gamma_\mu + f_2 v_\mu + f_3 v'_\mu \big]
  u(p,s)\,, \nn\\*
\langle \Lc(p',s')| \bar c\gamma_\nu\gamma_5 b |\Lb(p,s)\rangle
  &= \bar u(p',s') \big[ g_1 \gamma_\mu + g_2 v_\mu + g_3 v'_\mu \big] 
  \gamma_5\, u(p,s)\,,
\end{align}
where again $p = m_{\Lb}v$, $p' = m_{\Lc}v'$, 
and the spinors are normalized to $\bar u(p,s) u(p,s) = 2m$. 
The form factors beyond the SM are defined via~\cite{Bernlochner:2018bfn}
\begin{align}
\label{eqn:HQETffdefBSM}
\langle \Lc(p',s')| \bar c\, b |\Lb(p,s)\rangle
  &= h_S\, \bar u(p',s')\, u(p,s)\,, \nn\\*
\langle \Lc(p',s')| \bar c \gamma_5 b |\Lb(p,s)\rangle
  &= h_P\, \bar u(p',s')\, \gamma_5\, u(p,s)\,, \nn\\*
\langle \Lc(p',s')| \bar c\, \sigma_{\mu\nu}\, b |\Lb(p,s)\rangle
  &= \bar u(p',s') \big[ h_1\, \sigma_{\mu\nu}
  + i\, h_2 (v_\mu \gamma_\nu - v_\nu \gamma_\mu)
  + i\, h_3 (v'_\mu \gamma_\nu - v'_\nu \gamma_\mu) \nn\\*
  & \qquad\quad + i\, h_4 (v_\mu v'_\nu - v_\nu v'_\mu) \big] u(p,s)\,.
\end{align}
Matching Eqs.~\eqref{eqn:HQETffdefSM} and~\eqref{eqn:HQETffdefBSM} onto Eq.~\eqref{eqn:LbLcmatch}, 
the $\Lb \to \Lc$ form factors at $\mathcal{O}(\aS, 1/m_{c,b}^2)$ can be expressed in terms of the HQS functions as
\begin{subequations}
\label{eqn:ffmatchwfn}
\begin{align}
	\hat{h}_S& = 1 + \haS \Cs  + \sum_{Q = c,b} \eQ (\hK1{Q}-(w-1) \hK2{Q}) \nn \\
		& \qquad +\eb \ec (\hM1-2 \hM2 (w-1)+\hM3(w^2 -1)+\hM4 (w+2))\,, \\
	\hat{h}_P& = 1+ \haS \Cps + \sum_{Q = c,b} \eQ (\hK1{Q}-(w+1) \hK2{Q}) \nn \\
		& \qquad +\eb \ec (\hM1-2 \hM2 (w+1)+\hM3(w^2-1)+\hM4 (w-2))\,, \\
	\hat{f}_1& = 1+ \haS \Cv1 + \sum_{Q = c,b} \eQ (\hK1{Q}-(w+1) \hK2{Q}) \nn  \\
		& \qquad +\eb \ec (\hM1-2 \hM2 (w+1)+\hM3 (w^2-1)+\hM4 w)\, \\
	\hat{f}_2& = \haS \Cv2 + 2 \ec \hK2{c} + 2 \eb \ec (\hM2 -\hM3(w-1) -\hM4)\,, \\
	\hat{f}_3& =  \haS \Cv3 + 2 \eb \hK2{b} + 2 \eb \ec (\hM2 - \hM3(w-1) -\hM4)\,, \\
	\hat{g}_1& = 1+ \haS \Ca1 + \sum_{Q = c,b} \eQ (\hK1{Q}-(w-1) \hK2{Q})  \nn \\
		& \qquad + \eb \ec (\hM1-2 \hM2 (w-1)+\hM3(w^2-1)+\hM4 w)\,,  \\
	\hat{g}_2& =  \haS \Ca2 + 2 \ec \hK2{c} + 2 \eb \ec (\hM2-\hM3 (w+1)-\hM4)\,,  \\
	\hat{g}_3& =  \haS \Ca3 - 2 \eb \hK2{b} - 2 \eb \ec (\hM2-\hM3 (w+1)-\hM4)\,,  \\
	\hat{h}_1& = 1+ \haS \Ct1 + \sum_{Q = c,b} \eQ (\hK1{Q}-(w-1) \hK2{Q}) \nn \\
		& \qquad +\eb \ec (\hM1-2 \hM2 (w-1)+\hM3(w^2-1) +\hM4 (w-2))\,,  \\
	\hat{h}_2& = \haS \Ct2 + 2 \ec \hK2{c} +2 \eb \ec (\hM2-\hM3 (w+1)-\hM4)\,, \\
	\hat{h}_3& = \haS \Ct3 - 2 \eb \hK2{b} - 2 \eb \ec (\hM2-\hM3(w+1)-\hM4)\,, \\
	\hat{h}_4& = 4 \eb \ec \hM3\,.
\end{align}
\end{subequations}
Here we have defined, for convenience, 
\begin{equation}
	\label{eqn:hKcb}
	\hK{i}{Q} = \hK{i}1 + \eQ \hK{i}2\,,\qquad Q = c,b\,.
\end{equation}
We have included here the leading perturbative corrections in $\haS = \aS/\pi$~\cite{Neubert:1992qq};
explicit expressions for the $C_{\Gamma_i}$ functions are given in Ref.~\cite{Bernlochner:2017jka}.
The higher-order $\haS/m_{c,b}$ corrections have been previously computed in e.g. Refs.~\cite{Bernlochner:2018kxh, Bernlochner:2018bfn}, 
and are reproduced in Appendix \ref{app:radcrr}.

The six $b_i$ HQS functions, $i = 1,\ldots,6$, that encode the second-order power corrections in the notation of Ref.~\cite{Falk:1992ws} 
may be expressed in terms of the six functions $\hK{1,2}2$,  $\hM{1,2,3,4}$ via
\begin{align}
	\hat b_1/\LamB^2 & = \hK12 - (w-1) \hK22 		& \hat b_2/\LamB^2 & = 2 \hK22\,, \label{eqn:biKi}\\*
	\hat b_3/\LamB^2 & = \hM1 - 2\hM2(w-1) + \hM3(w^2-1) + \hM4 w		& \hat b_4/\LamB^2 & = 4\hM2\,, \nn \\*
	\hat b_5/\LamB^2 & = 2(\hM2 -  \hM3(w-1) - \hM4) 	& \hat b_6/\LamB^2 & = 2(\hM2 - \hM3 (w+1) - \hM4)\,, \nn
\end{align}
noting the $\hat{b}_i$ are dimensionful.
In Refs.~\cite{Bernlochner:2018kxh, Bernlochner:2018bfn}, which worked only to $\mathcal{O}(1/m_c^2)$,
the two independent subsubleading HQS functions entering at that order were expressed in terms of $\hat{b}_{1,2}$,
as done in Ref.~\cite{Falk:1992ws}, rather than in terms of $\hK{1,2}{2}$.

\subsection{Summary of constrained form factors}
\label{sec:summff}
The full derivation of the first and second-order power corrections to the $\Lb \to \Lc$ form factors at $\mathcal{O}(\theta^2)$ are provided in Appendix~\ref{app:ffderivation}.
As is well-known, at first order, the current corrections are fully determined by the leading Isgur-Wise function 
via the leading-order Schwinger-Dyson relation~\eqref{eqn:SDrelNLO} and equations of motion.
The first-order chromomagnetic corrections vanish, 
leaving a single subleading Isgur-Wise function associated with the $\mathcal{L}_1$ kinetic term, $\chi_1$.
The second-order current corrections are only partially determined 
by the leading Isgur-Wise function via the next-to-leading-order Schwinger-Dyson relation~\eqref{eqn:SDrelNNLO} and equations of motion.
At $\mathcal{O}(\theta^2)$ in the RCE, however, there is a single remaining Isgur-Wise function associated with second-order current corrections, $\varphi_1(w)$.
One finds that it is constrained at zero recoil such that $\hvph(1) = \lam1/(6\LamB^2)$.
Analyticity of the matrix elements ensures that the combination $[\hvph(w) - \hvph(1)]/(w-1)$ must be regular.
As done in Ref.~\cite{Bernlochner:2022ywh}, we define the \emph{quotient} with respect to $w=1$,
\begin{equation}
	\label{eqn:quotdef}
	\hvphq(w) \equiv [\hvph(w) - \hvph(1)]/(w-1)\,,
\end{equation}
and write the form factors explicitly in terms of this regular function.
By definition $\hvphq(1) = \hvph'(1)$, the gradient at zero recoil.
Finally, at $\mathcal{O}(\theta^2)$ in the RCE there is a single Isgur-Wise function 
associated with Lagrangian corrections from the $\mathcal{L}_2$ kinetic term, $\beta_1$.

The mass normalization condition~\eqref{eqn:massnormcon} requires that in the equal mass and zero-recoil limit the vector current matrix elements satisfy, to all orders,
\begin{equation}
	\label{eqn:f123zr}
	\big[f_1(1) + f_2(1) + f_3(1)\big]\Big|_{m_c = m_b} = 1\,.
\end{equation}
Thus, both the perturbative corrections and the power corrections must vanish in this limit order by order.
At first order, it immediately follows from Eq.~\eqref{eqn:ffmatchwfn} that $2\hK11(1) = 0$ and hence from Eq.~\eqref{eqn:Ki1}
\begin{equation}
	\label{eqn:chi1zr}
	\hat\chi_1(1) = 0\,.
\end{equation}
At second order, Eq.~\eqref{eqn:f123zr} implies that $2\hK12(1) + \hM3(1) - 3\hM4(1) = 0$. 
Applying Eqs.~\eqref{eqn:K22}, \eqref{eqn:Mi}, \eqref{eqn:K12} and the zero-recoil constraint~\eqref{eqn:vphzerorec} one finds
\begin{equation}
	\label{eqn:zrbeta}
	\hat\beta_1(1) = \frac{\lam1}{4\LamB^2}\,.
\end{equation}

The three Isgur-Wise functions $\beta_1$, $\chi_1$ and $\zeta$ are associated with the same HQET amplitude, $\Ub^{v'} \Gamma\, \Us^v$,
that conserves heavy quark spin symmetry.
Therefore within the form factors, these three Isgur-Wise functions always enter in the same linear combination.
Defining $\beta_1^\quot$ in the same manner as in Eq.~\eqref{eqn:quotdef}, 
we may reabsorb $\beta_1^\quot$ and $\chi_1$ into $\zeta$~\cite{Bernlochner:2022ywh} (see also e.g. Refs.~\cite{Bernlochner:2018bfn,Bernlochner:2018kxh,Manohar:2000dt}) 
via the replacement
\begin{equation}
	\label{eqn:replNNLO}
	\zeta + 2(\ec + \eb)\chi_1 + 2(\ec^2 + \eb^2)(w-1)\beta^\quot_1 \to \zeta\,.
\end{equation}
Although the reabsorption of $\chi_1$ introduces additional spurious $\mathcal{O}(1/m_{c,b}^2)$ terms, 
these terms are $\mathcal{O}(\theta^3)$ in the RCE or higher, and thus may be neglected,\footnote{In the full HQ expansion 
these induced $\mathcal{O}(1/m_{c,b}^2)$ terms cancel against $\chi_1$ terms arising from second-order Schwinger-Dyson relations,
such that $\chi_1$ can also be consistently reabsorbed per Eq.~\eqref{eqn:replNNLO} even at second order in the full power expansion.}
while induced $\mathcal{O}(\aS/m_{c,b})$ terms cancel against $\mathcal{O}(\aS/m_{c,b})$ perturbative corrections involving $\chi_1$.
Thus, applying the RCE at $\mathcal{O}(\theta^2)$, only a single subsubleading Isgur-Wise function $\varphi_1^\quot(w)$ enters the HQ expansion at $\mathcal{O}(1/m_{c,b}^2)$ after redefinitions. 
One finds finally (see Appendix~\ref{app:ffderivation})
\begin{align}
	\hK11 & = 0\,, & \hK21 &= -\frac{1}{w+1}\,,\nn\\*
	\hK12 & = \frac{\lam1}{2\LamB^2}\,, & \hK22 &= \frac{\lam1}{3\LamB^2} + 2 (w-1) \hvphq \,, \label{eqn:hKexps}
\end{align}
and
\begin{align}
	\hM3 & = \frac{1}{w+1}\bigg[ (w-2)\frac{\lam1}{6\LamB^2} + (w^2+2)\hvphq - \frac{w-2}{2(w+1)} \bigg]\,,\nn\\*
	\hM4 & = (3-w)\frac{\lam1}{6\LamB^2}  - w(w-1)\hvphq + \frac{w-1}{2(w+1)}\,, \label{eqn:hMexps}
\end{align}
with $\hM1 = \hM2 = 0$ as expected. 
In these results we have applied Eq.~\eqref{eqn:phpdef} to Eqs.~\eqref{eqn:K22} and~\eqref{eqn:Mi},
in order to express $\hK22$ and $\hM{3,4}$ in terms of the quotient $\hvphq$,
and applied Eq.~\eqref{eqn:zrbeta} to Eq.~\eqref{eqn:K12} to express $\hK12$ in terms of $\hat\beta_1^\quot$, 
which is then reabsorbed per Eq.~\eqref{eqn:replNNLO}.

In Table~\ref{tab:expansions} we summarize the Isgur-Wise functions entering up to and including $\mathcal{O}(1/m_{c,b}^2)$
in the full HQ expansion compared to $\mathcal{O}(\theta^2)$ in the RCE. 
In the former, six Isgur-Wise functions parametrize the six second-order HQS functions $\hK{1,2}{2}$ and $\hM{1,2,3,4}$.
Restricting to $\mathcal{O}(1/m_c^2)$ only, three Isgur-Wise functions---two full functions and a zero-recoil parameter---parametrize 
the two remaining second-order HQS functions $\hK{1,2}{2}$.
Thus at both $\mathcal{O}(1/m_c^2)$ and $\mathcal{O}(1/m_{c,b}^2)$ the second-order HQ expansion is non predictive,
in the sense that there are at least as many Isgur-Wise functions as there are HQS functions.
Because of this, in the BLRS approach of Refs.~\cite{Bernlochner:2018kxh, Bernlochner:2018bfn},
the two $\mathcal{O}(1/m_c^2)$ independent subsubleading HQS functions $\hat{b}_{1,2}$ were simply treated as independent functions.
This is to be compared to the constrained and predictive structure of the $\mathcal{O}(\theta^2)$ RCE-based description of the second-order power corrections,
for which a single subsubleading Isgur-Wise function $\hvph$ (plus hadron mass parameters) describe all six HQS functions at $\mathcal{O}(1/m_{c,b}^2)$.

\begin{table}[t]
\begin{tabular}{cc|c|c|cc}
\hline\hline
\multicolumn{2}{c|}{Expansions}  &  $1/m_{c,b}^0$  &  $1/m_{c,b}$  &  $1/m_c^2$ only  &  $1/m_{c,b}^2$  \\
\hline\hline
\multirow{1}{*}{HQET}  &  Functions &  $\zeta(w)$  &  $1$ 
	& \makecell{$\varphi_{0,1}^\quot(w)$, $e_3(1)$} & \makecell{$\varphi_{0,1}^\quot(w)$, $e_3(w)$,\\[-4pt] $d_{2,3}(w)$, $d_1(1)$   }   \\ 
	\hline
  \multirow{2}{*}{BLRS~\cite{Bernlochner:2018bfn,Bernlochner:2018kxh}}  &  Functions &  $\zeta(w)$  &  $1$ & $\hat{b}_{1,2}(w)$ & \nax \\
  &  Parameters  & $ \zeta'$, $\zeta''$  &  $\LamB$  & $\hat{b}_{1,2}(1)$ & \nax \\ 
  \multirow{2}{*}{RCE}  &  Functions &  $\zeta(w)$  &  $1$ & $\varphi_1^\quot(w)$ & $\varphi_1^\quot(w)$ \\
  &  Parameters  & $ \zeta'$, $\zeta''$, $\zeta'''$  &  $\LamB$  &  $\lam1$, $\hvphp$, $\hvphpp$ & $\lam1$, $\hvphp$, $\hvphpp$ \\  \hline
\hline\hline
\end{tabular}
\caption{Isgur-Wise functions order by order in the HQ expansion,
as they arise in the full HQ expansion~\cite{Falk:1992ws}, 
in the BLRS parameterization~\cite{Bernlochner:2018bfn,Bernlochner:2018kxh}, 
and in the RCE in this work,
after redefinitions and reabsorption into lower-order Isgur-Wise functions.
Also shown are the parameter (super)sets for each parametrization, not including mass scheme parameters.
The subsubleading Isgur-Wise functions $d_{2,3}$ arise from $\sim \mathcal{L}_1 \mathcal{L}'_1$ Lagrangian corrections involving a double operator product, 
while $e_3$ arises from mixed corrections $\sim \mathcal{L}_1\mJ_1$ (see Ref.~\cite{Falk:1992ws}).
At $\mathcal{O}(1/m_{c}^2)$ only, several of these can be further reabsorbed into lower-order Isgur-Wise functions,
up to the zero-recoil contribution $e_3(1)$.}
\label{tab:expansions}
\end{table}

\subsection{$1S$ scheme and numerical inputs}
We use the $1S$ scheme~\cite{Hoang:1998ng, Hoang:1998hm, Hoang:1999ye} 
to achieve cancellation of leading renormalon ambiguities from the mass parameter $\LamB$ 
against those in the factorially-growing coefficients of the $\aS$ perturbative power series~\cite{Bigi:1994em, Beneke:1994sw}.
In this scheme, $\mbS$ is defined as half of the perturbatively computed $\Upsilon(1S)$ mass,
such that the pole mass 
\begin{equation}
	\label{eqn:mb1S}
	m_b(\mbS) \simeq \mbS(1 +2\aS(m_b)^2/9 + \ldots)\,.
\end{equation}
When one computes just the leading $n_f$-dependence at high orders~\cite{Beneke:1994bc, Neubert:1994wq, Luke:1994xd}, 
the splitting of the bottom and charm quark pole mass $\dmbc \equiv m_b - m_c$ 
has a renormalon ambiguity only at third order in the heavy quark expansion.
Thus, as we are working at second order in the HQ expansion, we fix $m_c=m_b - \dmbc$.
The renormalization group evolution $\aS(\mu) = \aS(\aS(m_Z),m_Z; \mu)$
computed at four-loop order~\cite{Czakon:2004bu,VANRITBERGEN1997379} determines $\aS(m_b) \simeq 0.215$.

Because, however, $\mbS$ and $\dmbc$ are extracted numerically from fits to measurements of inclusive spectra at $\mathcal{O}(1/m_Q^3)$~\cite{Bauer:2002sh,Bauer:2004ve,Ligeti:2014kia},
third-order terms should be similarly retained numerically in the expansion of the hadron mass, 
even though we otherwise formally work to second order.
Defining $m_c(\mbS) \equiv m_b(\mbS) - \dmbc$, the mass expansion~\eqref{eqn:HmassHQE} then becomes
\begin{equation}
	\label{eqn:HmassHQE3} 
	m_{\Lambda_Q}  \simeq m_Q(\mbS) + \LamB - \frac{\lam1}{2m_Q(\mbS)} + \frac{\rh1}{4 [m_Q(\mbS)]^2}\,.
\end{equation}
Here at $\mathcal{O}(\theta^2)$ the third-order correction in the mass expansion is proportional to the parameter $\rh1$, 
in parallel to the expansions in Ref.~\cite{Gremm:1996df} for meson masses.
The mass expansion~\eqref{eqn:HmassHQE3} for $m_{\Lb}$ and $m_{\Lc}$ may be solved simultaneously to yield
\begin{align}
	\label{eqn:1Slams}
	\LamB & = \frac{m_b(\mbS) m_{\Lb} - m_c(\mbS)m_{\Lc}} {\dmbc} - \big[m_b(\mbS)  + m_c(\mbS)\big] + \frac{\rh1}{4m_b(\mbS) m_c(\mbS)}\,, \,, \nn \\
	\lam1 & = \frac{2 m_b(\mbS)m_c(\mbS)}{\dmbc}\big[ m_{\Lb} - m_{\Lc} - \dmbc\big] + \frac{\rh1\big[m_b(\mbS)  + m_c(\mbS)\big]}{2m_b(\mbS) m_c(\mbS)} \,,
\end{align}
so that $\LamB$ and $\lam1$ are parametrized in terms of $\mbS$, $\dmbc$, and $\rh1$.

Using the numerical inputs
\begin{equation}
	\label{eqn:mbSdmbc}
	\mbS = (4.71\pm 0.05)\,\GeV\,,\qquad \dmbc = (3.40\pm0.02)\,\GeV\,.
\end{equation}
one finds, keeping $\rh1$ dependence explicit, $\LamB = (0.88\pm0.05 + 0.04 [\rh1/\GeV^3])\,\GeV$ and $\lam1 = (-0.24\pm0.07 + 0.49[\rh1/\GeV^3])\,\GeV^2$.
As in Ref.~\cite{Bernlochner:2022ywh} we choose the HQET to QCD matching scale $\mubc = \sqrt{m_bm_c} \simeq 2.5\,\GeV$,
such that $\aS(\mubc) \simeq 0.27$, 
again via the renormalization group evolution $\aS(\mu) = \aS(\aS(m_Z),m_Z; \mu)$
computed at four-loop order~\cite{Czakon:2004bu,VANRITBERGEN1997379}.
As fit inputs we use the values in Eq.~\eqref{eqn:mbSdmbc} as well as
\begin{equation}
	\label{eqn:rh1val}
	\rh1 = (-0.1 \pm 0.2)\,\GeV^3\,,
\end{equation}
which corresponds to the range $\lam1 = (-0.3 \pm 0.1)\,\GeV^2$. 
This is roughly commensurate with the second-order parameter in the expansion of the spin-averaged $B^{(*)}$ and $D^{(*)}$ masses from fits to inclusive spectra~\cite{Ligeti:2014kia}.
The fit input values in Eqs.~\eqref{eqn:mbSdmbc} and Eq.~\eqref{eqn:rh1val} should not be confused 
with the central values and uncertainties recovered from any particular fit.

At first order in the HQ expansion, 
the $1S$ scheme is implemented by replacing the pole mass $m_b(\mbS)$ by $\mbS$ everywhere in the power corrections,
because all the first-order current corrections are fixed proportional to the leading Isgur-Wise function by the Schwinger-Dyson relation~\eqref{eqn:SDrelNLO},
and because there are no (unabsorbed) Lagrangian corrections.
Thus at first order, one may equivalently define and use an effective HQ expansion parameter
\begin{equation}
	\varepsilon^0_Q \equiv \frac{\LamB^{1S}}{2m_Q^{1S}} = \frac{1}{2m_Q^{1S}}\bigg[\frac{\mbS m_{\Lb} - \mcS m_{\Lc}} {\dmbc} - \big[\mbS  + \mcS\big] + \frac{\rh1}{4\mbS \mcS}\bigg]\,,
\end{equation}
in which $\mcS = \mbS - \dmbc$, 
while at second order one uses the full form of $\varepsilon_Q$ computed with the pole mass~\eqref{eqn:mb1S}.

\subsection{Parametrization of Isgur-Wise functions}
\label{sec:IWparam}
As done for the BLRS parametrization~\cite{Bernlochner:2018bfn,Bernlochner:2018kxh}, 
we use a simple $(w-1)$ expansion of the leading Isgur-Wise function,
such that
\begin{equation} 
	\label{eq:iw_w}
	\zeta(w) = 1 + \zeta' \, (w-1) + \frac{1}{2}\zeta''\, (w-1)^2 +  \frac{1}{6}\zeta'''\, (w-1)^3 \ldots\,,
\end{equation}
with the zero-recoil slope $\zeta' = \zeta'(1)$, curvature $\zeta'' = \zeta''(1)$, and $\zeta''' = \zeta'''(1)$.
We expand up to cubic order so that we may compare various RCE and BLRS fits with differing numbers of leading and subsubleading parameters,
while keeping the total number of free fit parameters fixed. 

One may instead contemplate expansion of $\zeta(w)$ with respect to the conformal coordinate 
\begin{equation}
	\label{eqn:zwtrans}
	z(w,w_0) = \frac{\sqrtb{w - w_+}- \sqrtb{w_0 - w_+}}{\sqrtb{w - w_+} + \sqrtb{w_0 - w_+}}\,,
\end{equation}
in which the branch point $w_+$ is formally determined by the invariant mass of the lightest hadron pair $\Hb \Hbar_c$---i.e. $q^2 = (m_{\Hb} + m_{\Hc})^2$---whose 
quantum numbers match onto the current in question, that is
\begin{equation}
	w_+ = -1 + \frac{(m_{\Lb} + m_{\Lc})^2 - (m_{\Hb} + m_{\Hc})^2}{2 m_{\Lb} m_{\Lc}}\,.
\end{equation}
While in principle the lightest $\Hb\Hbar_c$ pair differs for different currents, leading to different $z$, 
as in Ref.~\cite{Bernlochner:2022ywh} we choose the branch point universally to be determined by $\Hb= B$ and $\Hc = D$, 
in order that the leading-order Isgur-Wise function remains universal.
Formal differences in the branch points for different currents are encoded in higher-order corrections in the HQ expansion.

The transformation~\eqref{eqn:zwtrans} maps the pair-production threshold region $w \le w_+$ (i.e. $q^2 \ge (m_{\Hb} + m_{\Hc})^2$) 
to the unit circle $|z| = 1$ centered at $w= w_0$.
Defining $a_{0+}^2  \equiv (w_0 -w_+)/2$, then the leading Isgur-Wise function is expanded to cubic order as
\begin{multline} 
	\label{eq:iw_z}
	\frac{\zeta(w)}{\zeta(w_0)} =  1 - 8 a_{0+}^2 \rho^2\, z(w,w_0) + 16\big(2 c_0 a_{0+}^4 - \rho^2 a_{0+}^2\big)\,z(w,w_0)^2 \\ 
	+ \big(256 a_{0+}^6 d_0/3 + 128 a_{0+}^4 c_0 - 24\rho^2 a_{0+}^2\big)\, z(w,w_0)^3 + \ldots.
\end{multline}
By construction the slope at $w=w_0$,  $\zeta'(w_0)/\zeta(w_0) = -\rho^2$, 
the curvature $\zeta''(w_0)/\zeta(w_0) = c_0$, and $\zeta'''(w_0)/\zeta(w_0) = d_0$.
For the zero-recoil choice $w_0=1$, the conformal parameter is written $z(w,1)=z$, 
and one has simply $\zeta' = -\rho^2$, $\zeta'' = c_0$ and $\zeta''' = d_0$.
One may instead consider the optimized choice $z(w,w_*)=z_*$ such that
\begin{equation} 
	w_0 = w_* \equiv w_+ +\big[(1+r^2 - 2rw_+)(1-w_+)/(2r)\big]^{1/2}\,, 
\end{equation}
with $r = m_{\Lc}/m_{\Lb}$, 
that minimizes $|z(w_{\text{max}},w_0) - z(1,w_0)|$ in the physical recoil range $1 \le w \le w_{\text{max}}$.
For this choice, 
$\zeta'(w_*)/\zeta(w_*) = -\rho_*^2$, $\zeta''(w_*)/\zeta(w_*)  = c_*$ and $\zeta'''(w_*)/\zeta(w_*)  = d_*$,
denoting the leading-order parameters with a ``$^*$'' subscript.

The quotient of the single remaining subsubleading Isgur-Wise function $\hvph$ is parametrized up to linear order
\begin{equation}
	\hvphq(w) = \hvphp + \frac{1}{2} \hvphpp (w -1)\ldots\,,
\end{equation}
in which $\hvphp = \hvphp(1)$ and $\hvphpp = \hvphpp(1)$.
Thus in total there are five free fit parameters that may be considered
\begin{equation}
	\label{eqn:paramset}
	\zeta'\,, \quad \zeta''\,, \quad \zeta'''\,, \quad \hvphp\,,  \quad \text{and}  \quad \hvphpp\,,
\end{equation}
(or equivalently $\rho^2$, $c_0$, and $d_0$, or $\rho_*^2$, $c_*$, and $d_*$ for the leading-order parameters)
along with the Gaussian-constrained $\mbS$, $\dmbc$ and $\rh1$ per Eqs~\eqref{eqn:mbSdmbc} and~\eqref{eqn:rh1val}.
   
In the BLRS parametrization~\cite{Bernlochner:2018kxh, Bernlochner:2018bfn}, 
the two subsubleading HQS functions $\hat{b}_{1,2}(w)$ were parametrized as two constants, $\hat{b}_{1,2}$.
With reference to Sec.~\ref{sec:IWparam}, 
the total set of BLRS fit parameters comprised $\zeta'$, $\zeta''$, $\hat{b}_1$, and $\hat{b}_2$,
along with the Gaussian-constrained $\mbS$ and $\dmbc$.
The various sets of parameters used here and in the BLRS parametrization are summarized in Table~\ref{tab:expansions},
not including the $1S$ scheme parameters.
Thus to directly compare RCE-based fits with BLRS fit results, 
in this work we shall only examine fit scenarios involving subsets of four of the five parameters~\eqref{eqn:paramset}.

\subsection{Zero-recoil SM predictions}\label{sec:zerorecoilSM}
LQCD calculations~\cite{Detmold:2015aaa} provide predictions for the six SM form factors at zero recoil and beyond 
(see details in Sec.~\ref{sec:data} below). 
In Table~\ref{tab:lqcdzerorec} we show the corresponding zero-recoil central values, uncertainties and correlations:
$f_1(1)$, $f_2(1)$, $f_3(1)$, and $g_1(1)$ exhibit the most precise predictions.

\begin{table}[tb]
\renewcommand*{\arraystretch}{0.9}
\newcolumntype{C}{ >{\raggedleft\arraybackslash $} m{1.9cm} <{$}}
\newcolumntype{D}{ >{\centering\arraybackslash $} m{1cm} <{$}}
\begin{tabular}{DCCCCCC}
	\hline\hline
 & f_1(1) & f_2(1) & f_3(1) & g_1(1) & g_2(1) & g_3(1) \\
 \hline
 & 1.514(64) & -0.435(51) & -0.099(34) & 0.904(24) & -0.450(205) & 0.269(192) \\
\hline
f_1(1) & 1. & -0.595 & -0.340 & 0.165 & -0.101 & 0.059 \\
f_2(1) & \nax & 1. & -0.206 & -0.032 & 0.113 & -0.091 \\
f_3(1) & \nax & \nax & 1. & -0.040 & 0.007 & 0.038 \\
g_1(1) & \nax & \nax & \nax & 1. & -0.029 & 0.000 \\
g_2(1) & \nax & \nax & \nax & \nax & 1. & 0.674 \\
g_3(1) & \nax & \nax & \nax & \nax & \nax & 1. \\
\hline\hline
\end{tabular}
\caption{Central values, uncertainties and correlations of the LQCD predictions~\cite{Detmold:2015aaa} for the six SM form factors at zero recoil, 
obtained via the procedure outlined in Sec.~\ref{sec:data} below.}
\label{tab:lqcdzerorec}
\end{table}

At zero recoil and $\mathcal{O}(\aS, 1/m_{c,b}^2, \theta^2)$ in the RCE, the SM form factors have explicit form (noting $\hat{W}(1) = W(1)$ for each form factor)
\begin{subequations}
\label{eqn:FGzerorec}
\begin{align}
	f_1(1)_{\text{RC}} & = 1+\eb+\ec+\Cv1 \haS-(\eb-\ec)^2 \lamL[][6]\,, \\
	f_2(1)_{\text{RC}}  & = \Cv2 \haS -\ec + (\ec^2-\ec\eb) \lamL[2][3]\,, \\
	f_3(1)_{\text{RC}}  & = \Cv3 \haS -\eb+ (\eb^2 -\eb\ec) \lamL[2][3]\,,  \\
	g_1(1)_{\text{RC}}  & = 1+\Ca1 \haS+(3 \eb^2+2 \eb \ec+3 \ec^2) \lamL[][6]\,,  \\
	g_2(1)_{\text{RC}}  & = \Ca2 \haS - \ec - \ec \eb \bigg(\frac{1}{2}+6 \hvphp+\lamL[][3]\bigg)+\ec^2\lamL[2][3]\,,\\
	g_3(1)_{\text{RC}}  & =  \Ca3 \haS + \eb + \ec \eb \bigg(\frac{1}{2}+6 \hvphp+\lamL[][3]\bigg)-\eb^2\lamL[2][3]\,.
\end{align}
\end{subequations}
The additional $\mathcal{O}(\aS/m_{c,b})$ corrections can be read directly off Eqs.~\eqref{eqn:asmcorr}. 
Corresponding expressions for the scalar, pseudoscalar and tensor current form factors are provided in Appendix~\ref{app:NPzerorec}.
As derived above, within the RCE the second-order power corrections at zero recoil are fully determined by $\lam1$, 
except for $\mathcal{O}(\ec\eb)$ corrections in $g_{2,3}(1)$ and $h_{2,3,4}(1)$ that are also sensitive to $\hvphp$.
Note the first-order power corrections vanish in $g_1(1)$.
Analogously to the case for the axial-vector form factor $\mathcal{F}(1)$ in $\bar{B} \to D^*$ transitions,
the first-order prediction for $g_1(1)$
\begin{equation}
	\label{eqn:g1NLO}
	g_1(1)_{\text{NLO}} = 0.966(1)\,,
\end{equation}
is then extremely precisely determined, but is in tension with the LQCD predictions at $2.6\sigma$.
The size of this tension is at the several-percent level, corresponding to the expected size of second-order power corrections that may ameliorate it.

To this end, it is instructive to consider the numerical forms of Eqs.~\eqref{eqn:FGzerorec}, also including the $\mathcal{O}(\aS/m_{c,b})$ corrections. 
One finds
\begin{subequations}
\label{eqn:FGexpnum}
\begin{align}
	f_1(1) & = 1.518 + 0.028\big[\lam1/\GeV^2\big]\,, \\
	f_2(1) & = -0.379 + 0.039\big[\lam1/\GeV^2\big]\,, \\
	f_3(1) & = -0.116 - 0.029\big[\lam1/\GeV^2\big]\,,\\
	g_1(1) & = 0.966 + 0.092 \big[\lam1/\GeV^2\big]\,, \label{eqn:FGnum}\\
	g_2(1) & = -0.491 - 0.199 \hvphp  + \big[\lam1/\GeV^2\big] (0.046 -0.037 \hvphp)\,,\\
	g_3(1) & = 0.175 + 0.199 \hvphp + \big[\lam1/\GeV^2\big] (0.037\hvphp + 0.020)\,.
\end{align}
\end{subequations}
Here we have kept the dependence on $\lam1$ and the subsubleading Isgur-Wise parameter $\hvphp$ explicit:
the former arises from the $\lam1$ dependence in Eqs.~\eqref{eqn:FGzerorec}, 
as well as from $\eQ$ via the relations~\eqref{eqn:1Slams}, expressing $\LamB$ and $\rh1$ in terms of $\lam1$ itself.
For these numerical expressions, we have taken the central values $\mbS = 4.71\,\GeV$ and $\dmbc = 3.40\,\GeV$.
For the nominal range $\lam1 = (-0.3 \pm 0.1)\,\GeV^2$, the $\lam1$ term in Eq.~\eqref{eqn:FGnum} is $\sim -0.03$,
prospectively improving the agreement with the LQCD prediction ($g_1(1) = 0.904\pm 0.024$).

We also include the zero-recoil $\mathcal{O}(\aS^2)$ corrections, where known, 
to the vector and axial-vector currents~\cite{Czarnecki:1996gu, Czarnecki:1997cf, Franzkowski:1997vg}.
These amount to an additional shift
\begin{equation}
	\label{eqn:f1g1as2shift}
	\delta f_1(1) \simeq 0.509 C_F \haS^2 \simeq +0.005\,,\qquad \delta g_1(1) \simeq -0.944 C_F \haS^2 \simeq -0.009\,,
\end{equation}
where $C_F = (N_c^2 -1)/(2N_c)$. 	
In practice, in our fits in Sec.~\ref{sec:fits} we implement Eq.~\eqref{eqn:f1g1as2shift} via an overall shift in $f_1$ and $g_1$, such that
\begin{equation}
	\label{eqn:as2shift}
	\hat{f}_1 \to \hat{f}_1 + 0.509 C_F \haS^2\,,\qquad \hat{g}_1 \to \hat{g}_1 -0.944 C_F \haS^2\,.
\end{equation}

\begin{figure}[tb]
	\includegraphics[width = 10cm]{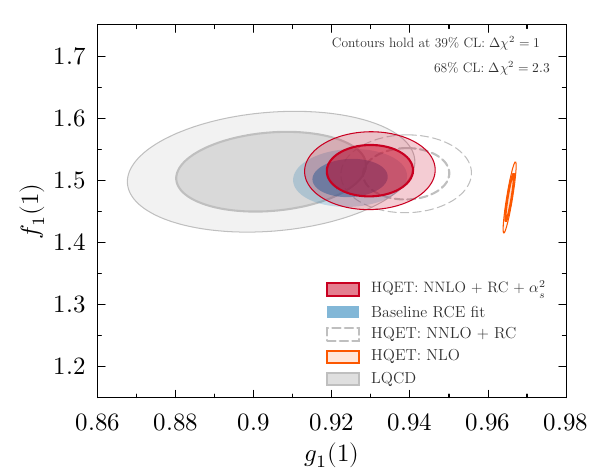}
	\caption{The RCE predictions for $g_1(1)$ and $f_1(1)$ at $\mathcal{O}(1/m_{c,b}^2, \aS/m_{c,b}, \aS^2, \theta^2)$ (red ellipse), 
	with uncertainties determined from the nominal ranges of $1S$ scheme fit inputs~\eqref{eqn:mbSdmbc} and \eqref{eqn:rh1val}:
	Dark and light shaded ellipses correspond to the $39\%$ and $68\%$ CLs.
	These are compared with the LQCD predictions (gray ellipses)
	and the HQET first-order predictions (orange ellipses), i.e. at $\mathcal{O}(1/m_{c,b}, \aS)$.
	Also shown are the RCE predictions without the $\mathcal{O}(\aS^2)$ corrections (gray dashed ellipses),
	and the results of the \basescen RCE fit scenario (blue ellipses) defined in Sec.~\ref{sec:fits} below.}
	\label{fig:g1f1}
\end{figure}

In Fig.~\ref{fig:g1f1} we show the CLs (red ellipse) in the $g_1(1)$--$f_1(1)$ plane
determined by the \emph{fit inputs}~\eqref{eqn:mbSdmbc} and \eqref{eqn:rh1val}, 
as well as the shifts~\eqref{eqn:f1g1as2shift} and the relations~\eqref{eqn:1Slams},
imposed on the $\mathcal{O}(1/m_{c,b}^2, \aS/m_{c,b}, \theta^2)$ form factor expressions.
The predicted central values and uncertainties are 
\begin{equation}
	g_1(1) = 0.930(11)\,, \qquad f_1(1)=1.515(41)\,,
\end{equation}
with a correlation of $0.02$.
This range is in agreement with the LQCD predictions (gray ellipse) at the $0.5\sigma$ level $(p =0.63)$.
If the $\mathcal{O}(\aS^2)$ correction in Eq.~\eqref{eqn:f1g1as2shift} is not included, 
the agreement is at the $0.8\sigma$ level $(p=0.42)$, indicated by the dashed gray ellipse.
By comparison, the first-order HQET CL (orange ellipse), 
whose small $g_1(1)$ uncertainty is determined by the first-order perturbative correction only as in Eq.~\eqref{eqn:g1NLO},
is approximately $2.2\sigma$ from the LQCD predictions.

If one instead examines the RCE-based zero-recoil predictions over the larger space of all form factors that are fully constrained at zero recoil, 
i.e. over the four-dimensional space spanned by $f_1(1)$, $f_2(1)$, $f_3(1)$, and $g_1(1)$, 
then one similarly finds agreement with LQCD predictions at the $0.3\sigma$ level $(p =0.78)$.
Without the $\mathcal{O}(\aS^2)$ corrections in Eq.~\eqref{eqn:f1g1as2shift}, this relaxes to the $0.4\sigma$ level ($p = 0.67$),
while the first-order HQET predictions are in tension with LQCD predictions at the $1.9\sigma$ level. 

\section{Fits}
\label{sec:fits}
\subsection{LHCb data and LQCD predictions}
\label{sec:data}
We consider fits to the LHCb measurement of the normalized differential spectrum $(1/\Gamma) \times  \d\Gamma(\Lb\to \Lc\mu\bar\nu)/\d q^2$~\cite{Aaij:2017svr} 
and to an LQCD determination of the six SM form factors~\cite{Detmold:2015aaa}. 
The measured LHCb spectrum is normalized to unity by the sum of the (arbitrary) reported yields, taking into account the reported correlations of Ref.~\cite{Aaij:2017svr}. 
Note that in the prior BLRS fit~\cite{Bernlochner:2018kxh} the small effects of these correlations were not incorporated: 
Their impact is of the order of $\sim 0.5\times 10^{-3}$ in $R(\Lc)$,
as well as an increase in the fit $\chi^2$ by $1.5$.\footnote{See footnote~\ref{ft:RLc}.} 
We incorporate this effect in all the BLRS fit results quoted in this work,
and unless otherwise specified, references hereafter to BLRS fits indicates fits including this minor modification 
versus the fits in Refs.~\cite{Bernlochner:2018kxh, Bernlochner:2018bfn}.
As we are analyzing a normalized spectrum, we remove one bin to avoid utilizing redundant degrees of freedom,
and therefore only fit six of the seven data points.

The LQCD results~\cite{Detmold:2015aaa} are published as fits to the BCL parametrization~\cite{Bourrely:2008za} with either 11 or 17 parameters.
Following the prescription in Ref.~\cite{Detmold:2015aaa}, we use the 11 parameter fit to obtain central values for synthetic LQCD data points sampled at three $q^2$ values---$q^2 = 1.0$, $5.6$, and $10.1\,\GeV^2$.
To generate uncertainties, we implement a modified version of the prescription outlined in Ref.~\cite{Detmold:2015aaa}, 
in order to account for the uncertainties from continuum and chiral extrapolation, 
truncation of the $z$ expansion, perturbative matching and other effects. 
We use the 17 parameter fit to determine a covariance matrix for all sampled $q^2$ values, 
and further include an additional uncorrelated uncertainty based on the difference between central values of the 11 and 17 parameter predictions. 

\subsection{Fit scenarios}
In order to compare with the BLRS fit results, 
which involved four free parameters $\zeta'$, $\zeta''$, $\hat{b}_{1,2}$, plus the Gaussian-constrained $\mbS$ and $\dmbc$,
we consider only four-parameter fit scenarios involving subsets of the five-parameter set~\eqref{eqn:paramset}.
We examine here three RCE-based fit scenarios, and consider two BLRS-based scenarios for comparison:
\begin{enumerate}[wide, labelwidth=0pt, labelindent=0pt, label = \textbf{(\roman*)}, noitemsep, topsep =0pt, itemjoin =\quad, series = fits]
	\item Using both LQCD predictions and LHCb data, with free parameters $\zeta'$, $\zeta''$, $\hvphp$, and $\hvphpp$ (i.e the leading-order cubic parameter $\zeta'''$ fixed to zero).
	We refer to this as our ``\basescen'' fit scenario.
	\item Using both LQCD predictions and LHCb data, with free parameters $\zeta'$, $\zeta''$, $\zeta'''$ and $\hvphp$ (i.e. with the subsubleading curvature parameter $\hvphpp$ fixed to zero). 
	We refer to this scenario as ``\basescen[\LOcubic]''.
	\item The \basescen scenario, but using only the LQCD results~\cite{Detmold:2015aaa} in the fit.
	We refer to this scenario as ``\lqcdscen''.	
	\item For comparison with the ``\basescen'' fit, the BLRS fit~\cite{Bernlochner:2018kxh} (as modified per Sec.~\ref{sec:data}) with free parameters $\zeta'$, $\zeta''$, $\hat{b}_1$, and $\hat{b}_2$. 
	\item For comparison with the ``\basescen[\LOcubic]'' fit, and noting $\hat{b}_2$ is compatible with zero in the (prior) BLRS fit (see Ref.~\cite{Bernlochner:2018kxh}), 
	we perform a BLRS fit with free parameters $\zeta'$, $\zeta''$, $\zeta'''$, and $\hat{b}_1$ (i.e. with $\hat{b}_2$ fixed to zero).
	This fit scenario is referred to as ``\blrsscen[\LOcubic]''.
\end{enumerate}
Alternate fits using the $z$ and $z_*$ expansions for the leading Isgur-Wise function,
per Sec.~\ref{sec:IWparam} and Eq.~\eqref{eq:iw_z}, yield near identical fit results for each fit scenario,
up to small improvements in fit quality, $\Delta \chi^2 \lesssim 1$.
We therefore do not consider $z$ and $z_*$-expansion-based fits further.

\subsection{Fit results}
The recovered fit parameters for each of the five fit scenarios are shown in Table~\ref{tab:fitres}.
The corresponding fit correlations for each fit scenario are provided in Appendix~\ref{sec:fitcorr}.
The slope and curvature of the leading Isgur-Wise function in the \basescen scenario is consistent with the BLRS fit. 
Noting the moderate anticorrelation of the subsubleading IW parameters $\hvphp$ and $\hvphpp$ 
($\rho = -0.51$, per Table~\ref{tab:fitcorrbasescen})
the fit for subsubleading two-parameter space is $\simeq1\sigma$ from zero ($p = 0.33$).
We see also that $\rh1$ has a relatively weak effect on the fit, as it is consistent with zero 
and only weakly correlated with the four free fit parameters.
Constraining $\rh1$ to be zero instead yields a near identical fit, with $\chi^2/\text{ndf} = 12.4/20$ and $\lam1 = -0.27(7)\,\GeV^2$.

\begin{table}[t]
\newcolumntype{C}{ >{\raggedleft\arraybackslash $} m{2.8cm} <{$}}
\newcolumntype{D}{ >{\centering\arraybackslash $} m{2.5cm} <{$}}
\renewcommand{\arraystretch}{0.9}
\resizebox{0.95\textwidth}{!}{\begin{tabular}{D|CCCCC}
\hline
\text{Scenario} 	 &\makecell[r]{\text{\basescen}}	 & \makecell[r]{\text{\basescen[\LOcubic]}}	 & \makecell[r]{\text{\lqcdscen}}	 & \makecell[r]{\text{BLRS}}	 & \makecell[r]{\text{BLRS\LOcubic}}  \\ 
 \text{Fit inputs} 	 & \makecell[r]{\text{LHCb+LQCD}}	 & \makecell[r]{\text{LHCb+LQCD}}	 & \makecell[r]{\text{LQCD}}	 & \makecell[r]{\text{LHCb+LQCD}}	 & \makecell[r]{\text{LHCb+LQCD}}  \\ 
 \hline\hline
\zeta'	 & -2.16 \pm 0.09	 & -2.74 \pm 0.24	 & -2.15 \pm 0.11	 & -2.06 \pm 0.08	 & -2.45 \pm 0.26  \\ 
\zeta''	 & 3.37 \pm 0.38	 & 11.40 \pm 3.04	 & 3.18 \pm 0.41	 & 3.31 \pm 0.37	 & 8.31 \pm 3.23  \\ 
\zeta'''	 & \nax	 & -37.40 \pm 14.50	 & \nax	 & \nax	 & -23.80 \pm 15.30  \\ 
\mbS~[\GeV]	 & 4.72 \pm 0.04	 & 4.71 \pm 0.04	 & 4.71 \pm 0.04	 & 4.73 \pm 0.05	 & 4.70 \pm 0.04  \\ 
\dmbc~[\GeV]	 & 3.41 \pm 0.02	 & 3.40 \pm 0.02	 & 3.41 \pm 0.02	 & 3.40 \pm 0.02	 & 3.40 \pm 0.02  \\ 
\hvphp	 & 0.09 \pm 0.38	 & -0.10 \pm 0.31	 & 0.04 \pm 0.33	 & \nax	 & \nax  \\ 
\hvphpp	 & -4.88 \pm 3.52	 & \nax	 & \nax	 & \nax	 & \nax  \\ 
\rh1~[\GeV^3]	 & -0.17 \pm 0.19	 & -0.15 \pm 0.19	 & -0.19 \pm 0.19	 & \nax	 & \nax  \\ 
\hat{b}_1~[\GeV^2]	 & \nax	 & \nax	 & \nax	 & -0.45 \pm 0.15	 & -0.36 \pm 0.15  \\ 
\hat{b}_2~[\GeV^2]	 & \nax	 & \nax	 & \nax	 & -0.25 \pm 0.39	 & \nax  \\ 
\hline
\chi^2 	 &11.8	 & 7.4	 & 3.7	 & 8.9	 & 7.0  \\ 
\text{ndf} 	 &20	 & 20	 & 14	 & 20	 & 20  \\ 
\lam1~[\GeV^2] 	 &-0.35 \pm 0.11	 & -0.33 \pm 0.11	 & -0.36 \pm 0.11	 & -0.24 \pm 0.07	 & -0.24 \pm 0.07  \\ 
\hline
R(\Lc) 	 & 0.3251 \pm 0.0037	 & 0.3214 \pm 0.0042	 & 0.3389 \pm 0.0102	 & 0.3233 \pm 0.0037	 & 0.3207 \pm 0.0041  \\ 
\text{Corr.} 	 & \text{Tab.~\ref{tab:fitcorrbasescen}}	 & \text{Tab.~\ref{tab:fitcorrbasescen[LOcubic]}}	 & \text{Tab.~\ref{tab:fitcorrlqcdscen}}	 & \text{Tab.~\ref{tab:fitcorrBLRS}}	 & \text{Tab.~\ref{tab:fitcorrblrsscen[LOcubic]}}  \\ 
\hline\hline
\end{tabular}}
\caption{Fit results for the three RCE fit scenarios plus the two comparison BLRS fit scenarios.
Also shown are the fit $\chi^2$, the degrees of freedom (ndf), and the recovered values for $\lam1$ and $R(\Lc)$,
and the corresponding tables containing the fit correlations.}
\label{tab:fitres}
\end{table}

The \basescen[\LOcubic] fit features a substantial improvement in fit quality versus the \basescen fit, 
and slightly better than BLRS. All three leading-order parameters are recovered significantly from zero,
while $\hvphp$ is consistent with zero but mildly correlated with the other free fit parameters,
as seen in Table~\ref{tab:fitcorrbasescen[LOcubic]}.
Thus a parametrization using the highly-constrained structure of RCE at second-order plus a constant $\hvphq(w) = \hvphp$ 
can be as compatible with data as BLRS, provided there is sufficient freedom in the leading order Isgur-Wise function.
The same behavior is seen for the \blrsscen[\LOcubic] fit scenario: $\hat{b}_2$, 
which is compatible with zero in the BLRS fit, 
can be fixed to zero and replaced by the leading-order cubic parameter $\zeta'''$, leading to a mild improvement in fit quality, while still compatible with the BLRS fit.
Thus the RCE-based fits do not appear to introduce biases compared to the performance of the BLRS fits.
Finally, the ``\lqcdscen'' fit features inflated uncertainties, but with otherwise qualitatively similar fit results,
just as was seen in the prior ``LQCD only'' fit for the BLRS parametrization~\cite{Bernlochner:2018kxh}.

\begin{figure}[t]
\includegraphics[width=0.49\textwidth]{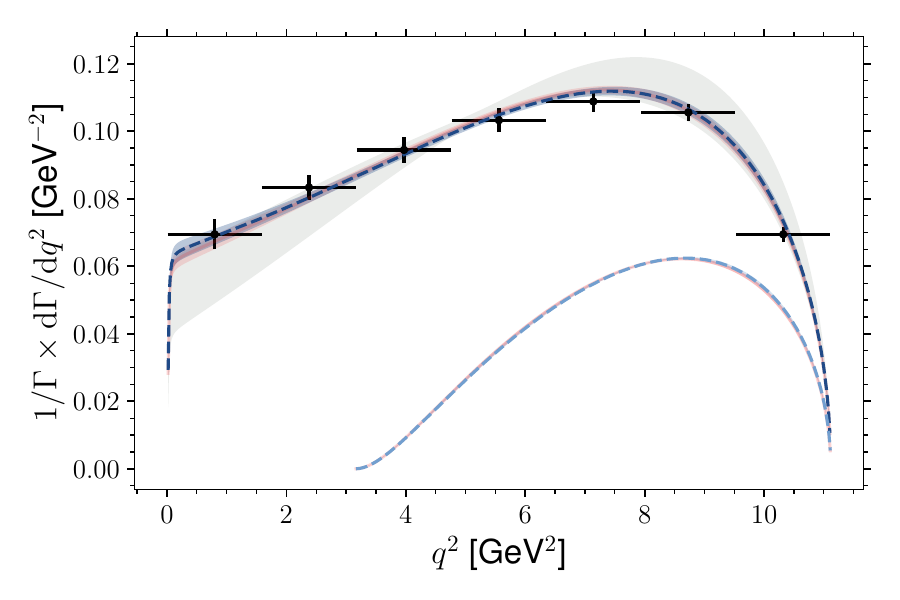}
\includegraphics[width=0.49\textwidth]{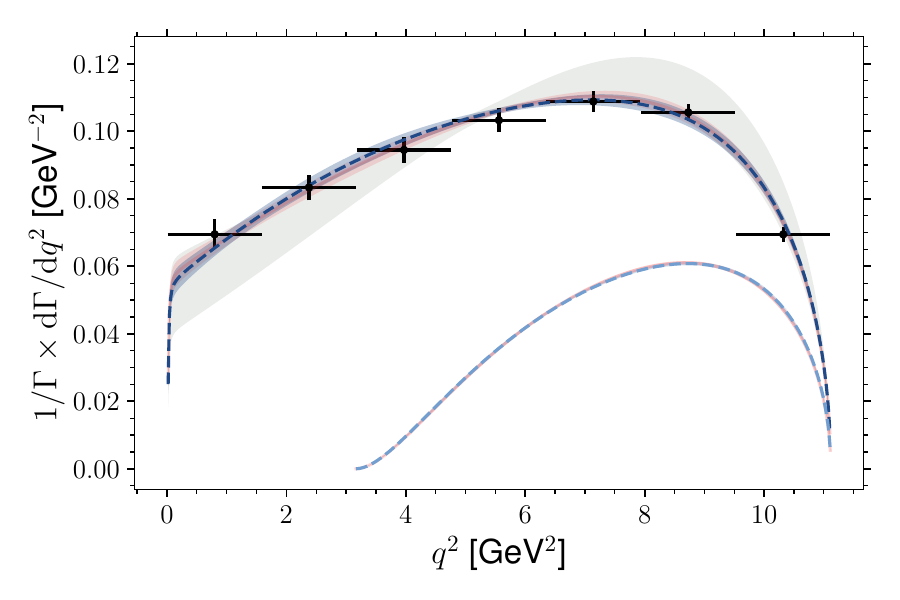}
\caption{Fit results for the normalized $\d\Gamma(\Lb \to \Lc \mu\nu)/\d q^2$ spectrum. 
Left: The 68\% CLs for the RCE-based \basescen fit results (blue bands) compared to the BLRS fit (red bands).
The data points correspond to the LHCb measurement~\cite{Aaij:2017svr}, while the gray bands show the 68\% CLs for the LQCD predictions. 
Also shown is the predicted spectrum for the $\Lb \to \Lc \tau \nu$ mode for the \basescen fit (light blue band), versus BLRS (light red band).
Right: The same but for the \basescen[\LOcubic] fit scenario compared to the \blrsscen[\LOcubic] fit scenario. See text for details.}
\label{fig:q2spec_fits}
\end{figure}

\begin{figure*}[t]
\includegraphics[width=0.45\textwidth]{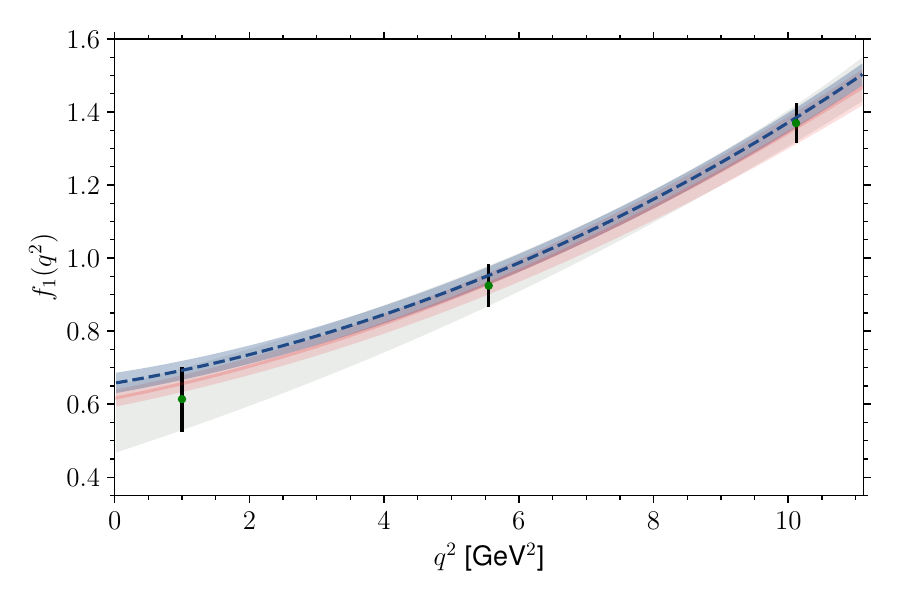}\hfil
\includegraphics[width=0.45\textwidth]{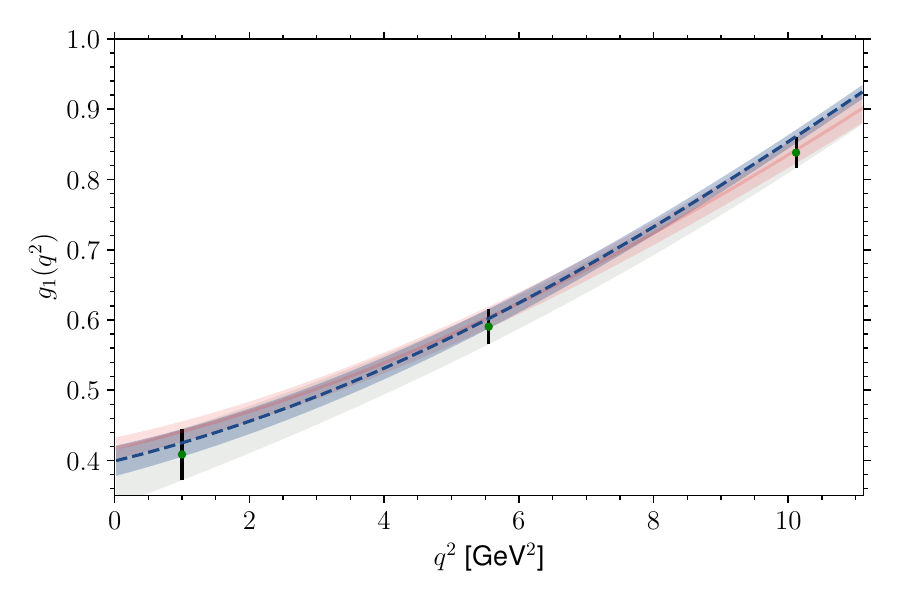}
\\[4pt]
\includegraphics[width=0.45\textwidth]{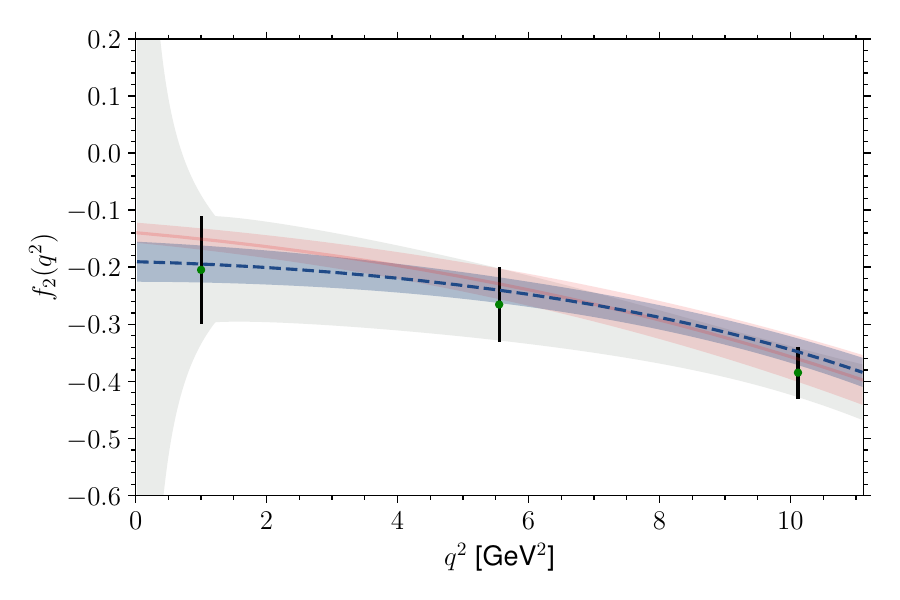}\hfil
\includegraphics[width=0.45\textwidth]{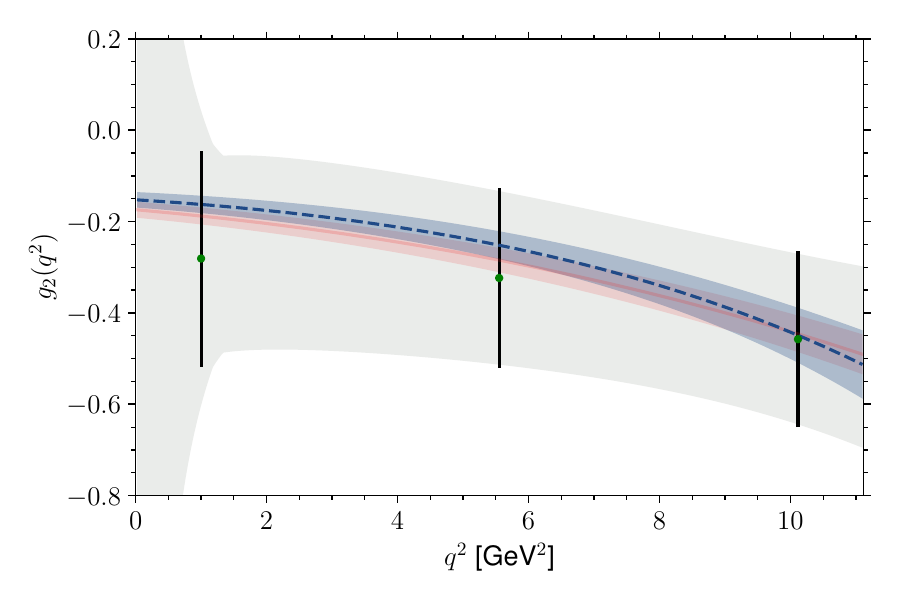}
\\[4pt]
\includegraphics[width=0.45\textwidth]{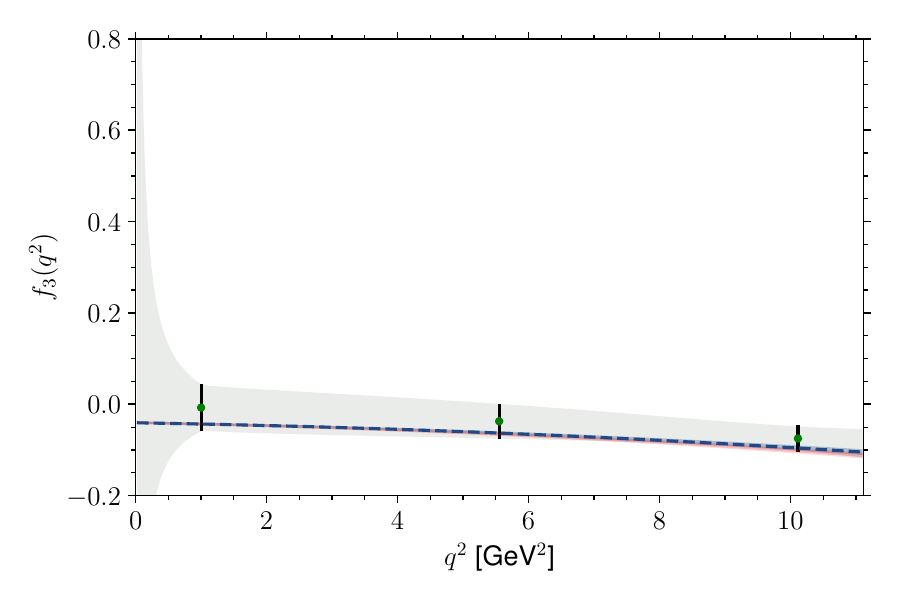}\hfil
\includegraphics[width=0.45\textwidth]{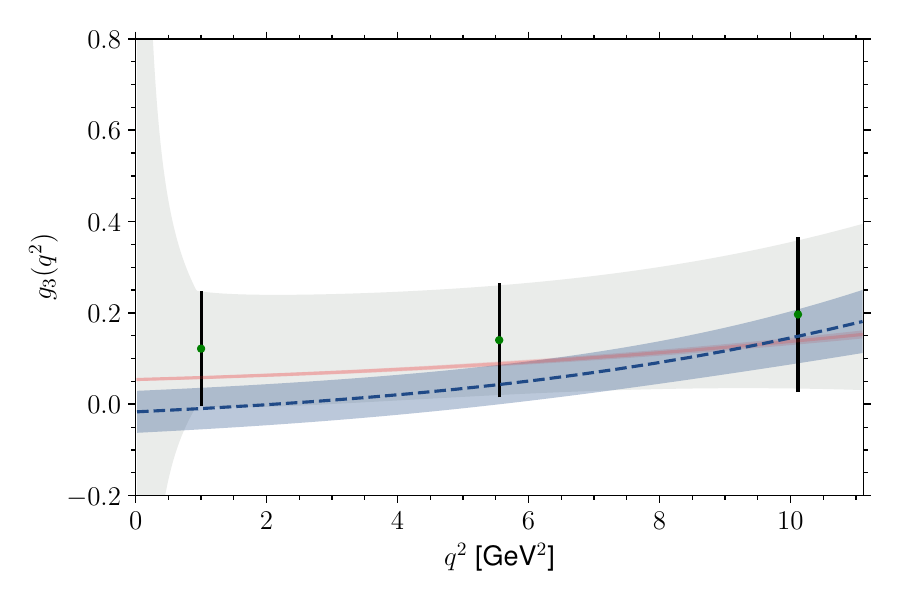}
\caption{Fit 68\% CLs for the \basescen RCE fit scenario for the six SM form factors (blue dashed bands), 
compared to the LQCD predictions (gray bands and data points)~\cite{Detmold:2015aaa}.
Also shown are the BLRS fit 68\% CLs (red solid line and bands; see also Figs~2 of Refs.~\cite{Bernlochner:2018kxh,Bernlochner:2018bfn}).}
\label{fig:ff_fits}
\end{figure*}

The $\Lb \to \Lc \ell \nu$ differential $q^2$ spectrum for the RCE-based \basescen fit scenario (68\% CLs, blue bands) is shown in Fig.~\ref{fig:q2spec_fits} (left),
compared to LHCb data (black points) and the LQCD predictions (gray bands), and the BLRS fit (red bands).
Also shown are the predicted spectra for the $\Lb \to \Lc \tau \nu$ mode. 
The resulting spectra are near identical for the \basescen and BLRS fits.
A similar result is seen in Fig.~\ref{fig:q2spec_fits} (right), comparing the \basescen[\LOcubic] and \blrsscen[\LOcubic] fit results. 
The noticeably higher curvature at lower $q^2$, enabled by the additional leading-order parameter, is mainly responsible for the improvement in fit quality.

Fit results for the six SM form factors are shown in Fig.~\ref{fig:ff_fits} (68\% CLs, blue bands),
compared to the LQCD predictions (gray bands) and synthetic data (data points),
as well as the results for the BLRS fits (red bands, see also Ref.~\cite{Bernlochner:2018kxh}).
As expected, the highly-constrained structure of $f_1(1)$ and $g_1(1)$ in the RCE leads to somewhat tighter predictions near maximal $q^2$
than for the BLRS fit, though the behavior across $q^2$ is otherwise very similar.
The zero-recoil fit CLs for $g_1(1)$ and $f_1(1)$ are shown by the blue ellipses in Fig.~\ref{fig:g1f1}.

The behavior and characteristic size of uncertainties in $f_2$ and $g_2$ are similar for the RCE \basescen and BLRS fit:
With reference to Eqs~\eqref{eqn:ffmatchwfn}, both of these form factors feature the same $2 \ec^2 \hK22 = \ec^2 \hat{b}_2/\LamB^2$ second-order term,
that dominates the other $\mathcal{O}(\ec\eb)$ corrections and generate the largest source of uncertainties. 
In both the RCE and BLRS parametrizations, one then expects similar fit uncertainties for these two form factors.

In BLRS both $f_3$ and $g_3$ feature tiny uncertainties.
However, in the RCE \basescen fit, only the uncertainties in $f_3$ are similarly small, 
while the uncertainties in $g_3$ are comparable to the other form factors.
The reason for this behavior in BLRS is that $f_3$ and $g_3$ feature only $\mathcal{O}(\eb^2)$ and $\mathcal{O}(\ec\eb)$ corrections, 
which were omitted in the BLRS parametrization entirely,
and thus there is no uncertainty in $f_3$ or $g_3$ arising from the second-order BLRS fit parameters $\hat{b}_{1,2}$.
The reason for the small uncertainties for $f_3$ versus $g_3$ in the RCE lies in the structure of the dominant second-order power corrections at $\mathcal{O}(\theta^2)$:
To see this, substituting the $\hK{i}{2}$ and $\hM{i}$ expressions in Eqs~\eqref{eqn:hKexps} and~\eqref{eqn:hMexps} into Eqs~\eqref{eqn:ffmatchwfn}, 
one finds that the dominant second-order power corrections in $\hat{f}_{3}$ and $\hat{g}_{3}$ are
\begin{subequations}
\begin{align}
	\delta \hat{f}_3 &= \ec\eb\bigg[\lamL[][3]\frac{w-5}{w+1} + 2\frac{(w-2)(w-1)}{w+1} \hvphq - 3 \frac{w-1}{(w+1)^2}\bigg]\,,\\
	\delta \hat{g}_3 &= \ec\eb\bigg[\lamL[][3] + 2(w+2) \hvphq + \frac{1}{w+1}\bigg]\,.
\end{align}
\end{subequations}
In these corrections, the $\hvphq$ terms are nominally the main source of fit uncertainties.	
The prefactor of the $\hvphq$ term in $\hat{f}_3$ happens to be heavily suppressed by the $w$-dependence, 
being at most $\sim 3\%$ of the size of the $\hvphq$ term in $\hat{g}_3$.
The latter term in turn happens to be of similar size to the $\hvphq$ terms in the other SM form factors, 
which arise from terms $\sim \ec^2 \hK{2}{2}$,
because of the enhancement by the $(w+2)/[2(w-1)]$ relative factor, that compensates the $\eb/\ec$ relative suppression.
Thus within the fit, the uncertainty in $f_3$ is greatly suppressed.

\begin{figure*}[t]
\includegraphics[width=0.45\textwidth]{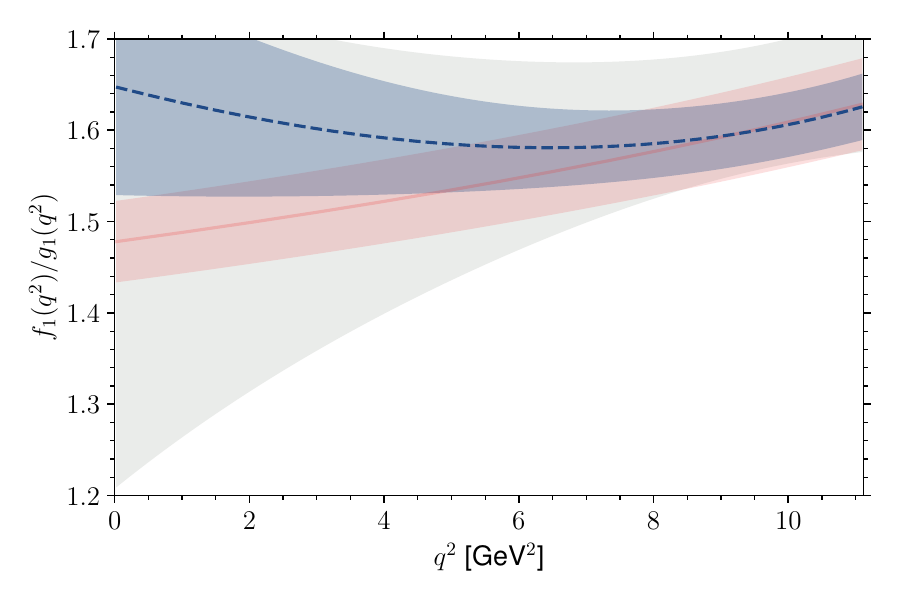}
\\[4pt]
\includegraphics[width=0.45\textwidth]{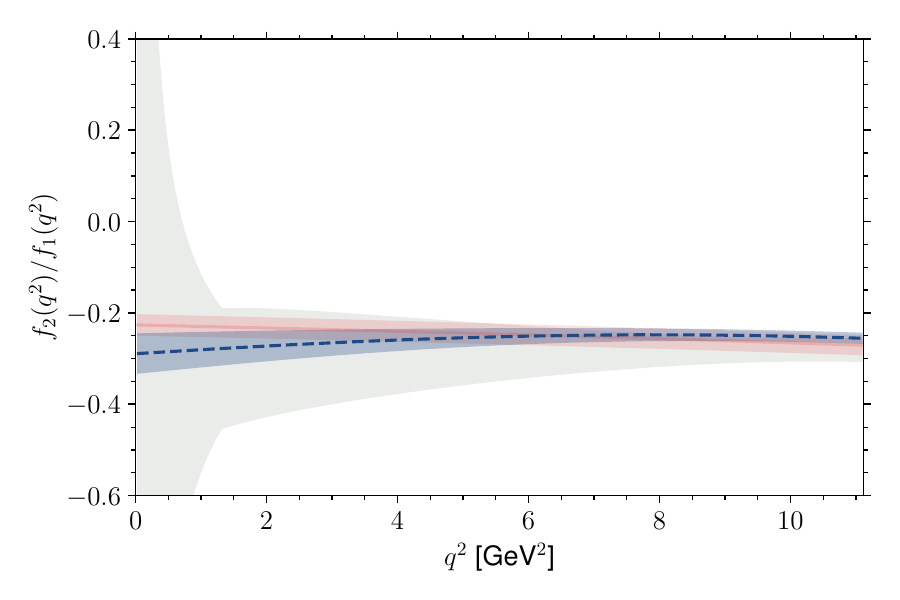}\hfil
\includegraphics[width=0.45\textwidth]{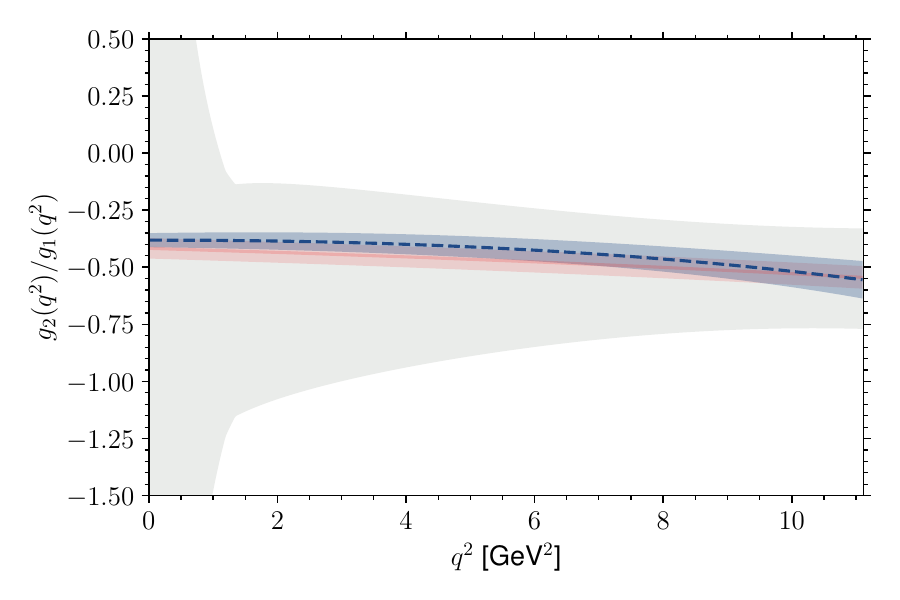}
\\[4pt]
\includegraphics[width=0.45\textwidth]{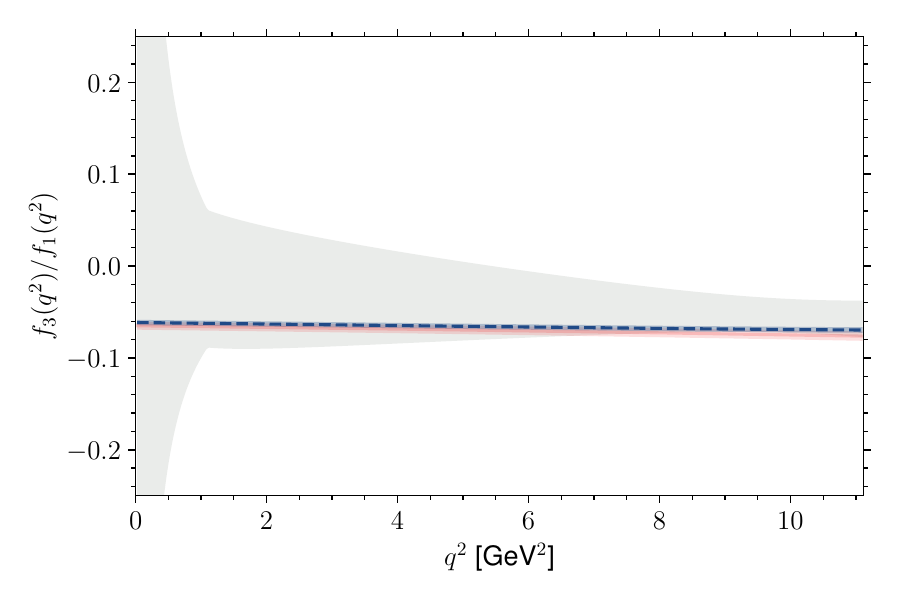}\hfil
\includegraphics[width=0.45\textwidth]{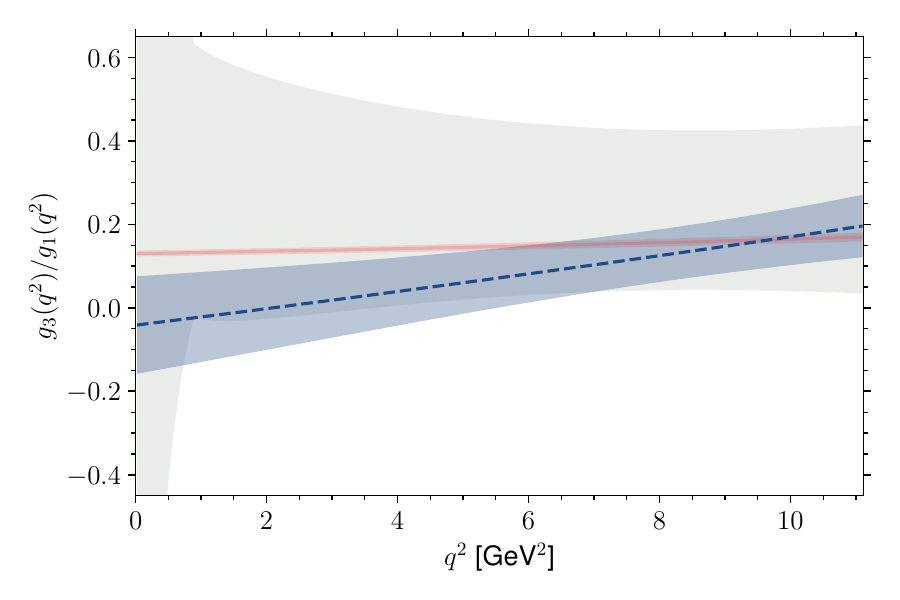}
\caption{Fit 68\% CLs for five ratios of the six SM form factors in the \basescen RCE-based fit scenario (blue dashed bands) 
compared the LQCD predictions (gray bands)~\cite{Detmold:2015aaa}.
The top row shows $f_1/g_1$, which is $\mathcal{O}(1)$ in the HQ expansion, 
while the left column shows $f_{2,3}/f_1$ and the right column shows $g_{2,3}/g_1$, all of which are expected to be $\mathcal{O}(\aS, \lqcd/m_{c,b})$. 
Also shown are the BLRS fit 68\% CLs (red bands); see also Fig.~3 of Ref.~\cite{Bernlochner:2018bfn}.}
\label{fig:ff_fit_ratios}
\end{figure*}

In order to factor out sensitivity to the leading Isgur-Wise function, 
in Fig.~\ref{fig:ff_fit_ratios} we also examine five form factor ratios:
$f_1/g_1$, which one expects should be leading order in the HQ expansion, as well as $f_{2,3}/f_1$, and $g_{2,3}/g_1$, 
that are expected to be non-zero only at first order and beyond in the HQ power expansion.
The CLs for the \basescen fit scenario (blue bands) are compatible with the LQCD predictions (gray bands) and these HQ power-counting expectations.
However, they do show some differences versus BLRS fits (red bands, see also Ref.~\cite{Bernlochner:2018bfn}).
In particular, differences in shape and uncertainty in $f_1/g_1$ are somewhat pronounced at low $q^2$, 
while the uncertainties in $g_3/g_1$ are substantially larger for similar reasons as discussed immediately above
(being careful to note one must include uncertainties from the second-order corrections entering $g_1$).

\begin{figure*}[t]
\includegraphics[width=0.45\textwidth]{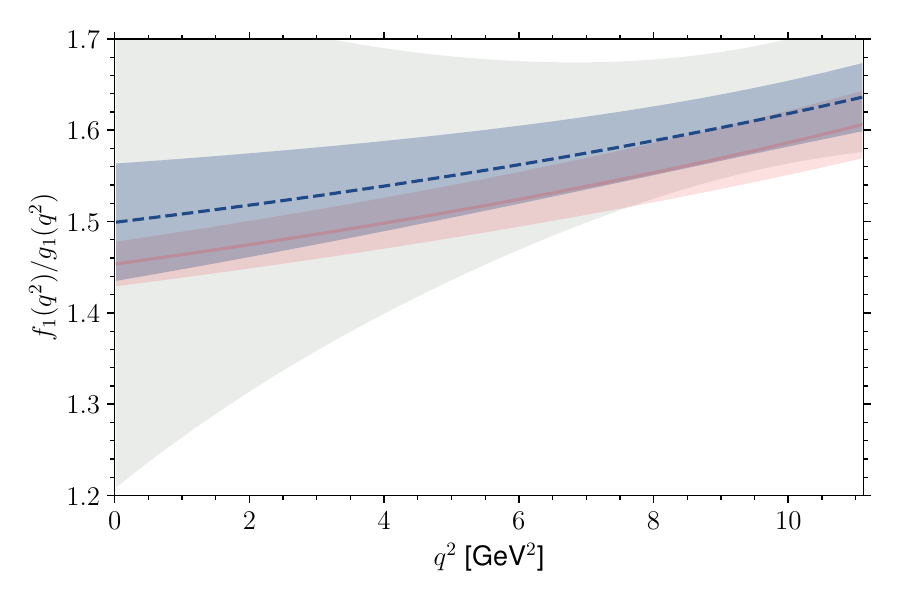}
\\[4pt]
\includegraphics[width=0.45\textwidth]{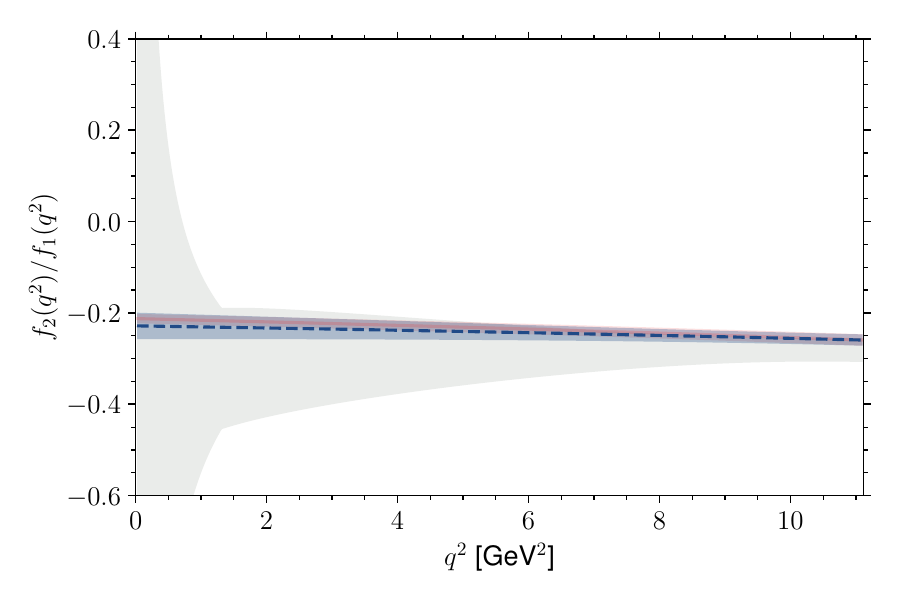}\hfil
\includegraphics[width=0.45\textwidth]{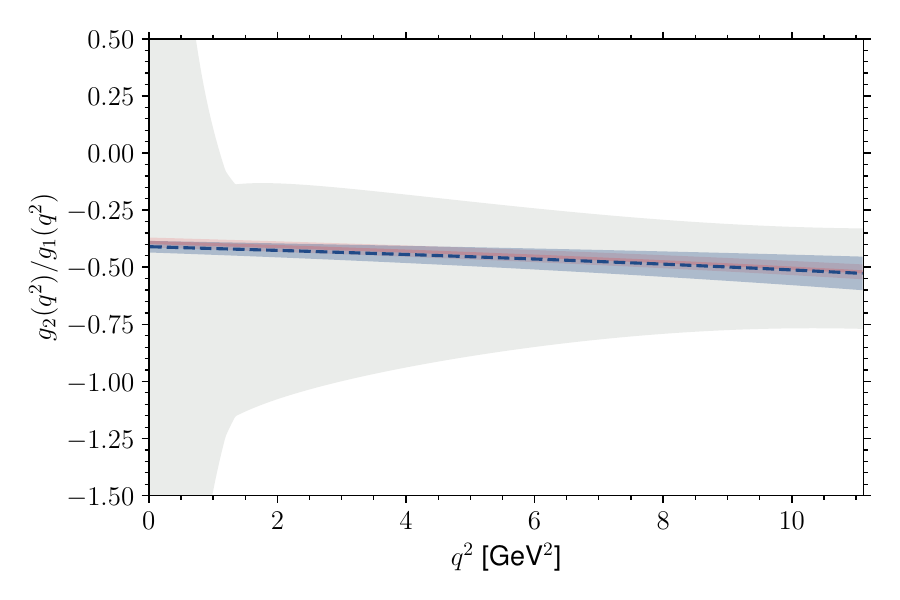}
\\[4pt]
\includegraphics[width=0.45\textwidth]{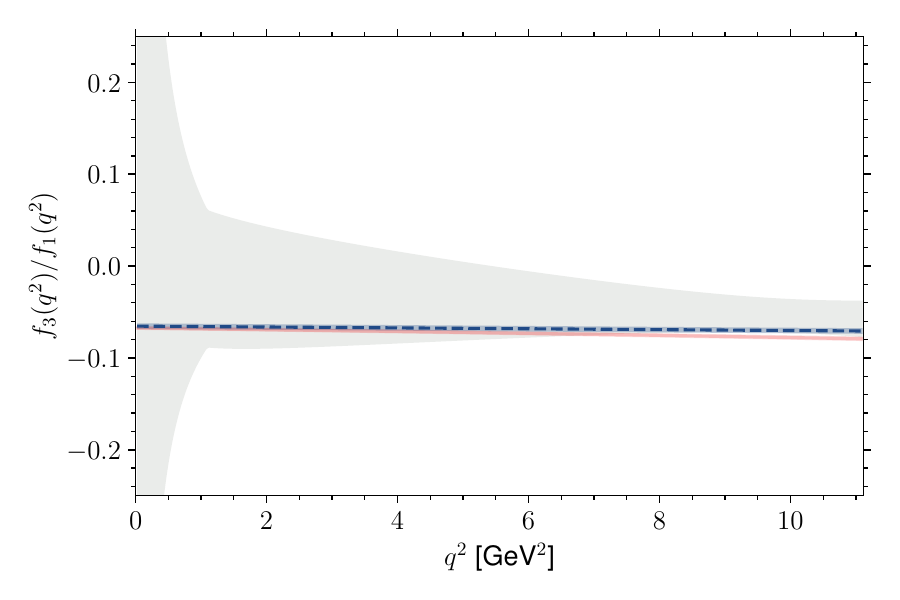}\hfil
\includegraphics[width=0.45\textwidth]{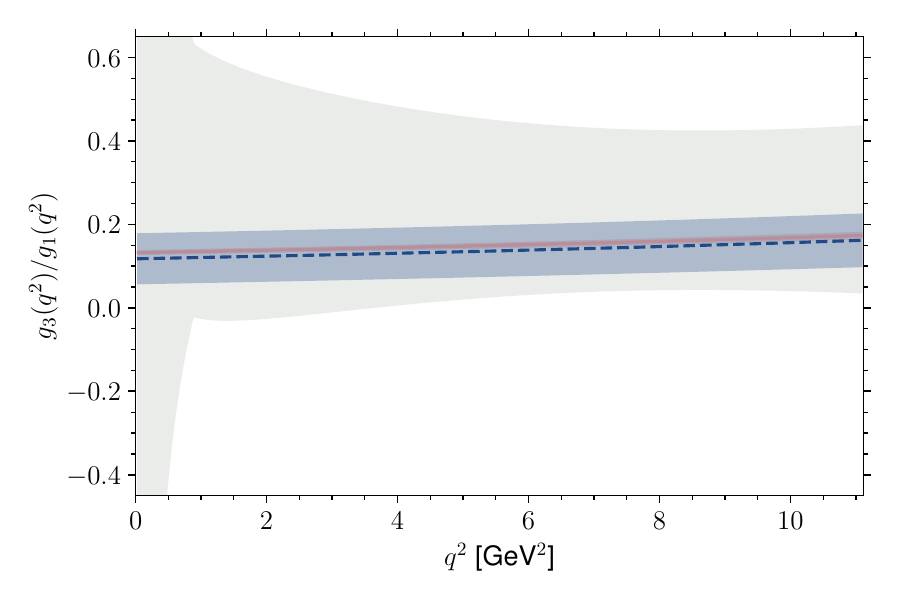}
\caption{Fit 68\% CLs for five ratios of the six SM form factors in the \basescen[\LOcubic] RCE-based fit scenario (blue dashed bands) 
compared the LQCD predictions (gray bands)~\cite{Detmold:2015aaa}.
The top row shows $f_1/g_1$, while the left column shows $f_{2,3}/f_1$ and the right column shows $g_{2,3}/g_1$. 
Also shown are the \blrsscen[\LOcubic]  fit 68\% CLs (red bands).}
\label{fig:ff_fit_ratios_LO3}
\end{figure*}

In Fig.~\ref{fig:ff_fit_ratios_LO3} we show the same set of ratios, comparing the predictions for the \basescen[\LOcubic] fit versus the \blrsscen[\LOcubic] scenario.
One sees that an RCE-based fit with a single, constant subsubleading parameter $\hvphp$ 
produces substantially similar results to an HQET-based fit with a single, constant subsubleading parameter $\hat{b}_1$---i.e.
fixing $\hat{b}_2$ to zero---even though, 
with reference to Eqs.~\eqref{eqn:biKi} and~\eqref{eqn:hKexps}, 
the zero-recoil predictions for $\hat{b}_2$ in the RCE are non-zero and tightly constrained.

Finally, the $R(\Lc)$ values for all three scenarios shown in Table~\ref{tab:fitres} are consistent with each other and the (prior) BLRS value at $<1\sigma$.
In Table~\ref{tab:RLcBFs} we show the corresponding partial widths $\Gamma[\Lb \to \Lc \mu \nu]/|V_{cb}|^2$ and $\Gamma[\Lb \to \Lc \tau \nu]/|V_{cb}|^2$ for the RCE baseline scenario,
as well as correlations with each other and $R(\Lc)$. 

\begin{table}[tb]
\renewcommand*{\arraystretch}{0.9}
\newcolumntype{C}{ >{\centering\arraybackslash $} m{3.5cm} <{$}}
\newcolumntype{D}{ >{\centering\arraybackslash $} m{2cm} <{$}}
\scalebox{0.85}{\begin{tabular}{DCC}
\hline\hline
R(\Lc) & \makecell[c]{\Gamma[\Lb \to \Lc \mu \nu]/|V_{cb}|^2 \\[-4pt] [\GeV]} & \makecell[c]{\Gamma[\Lb \to \Lc \tau \nu]/|V_{cb}|^2 \\[-4pt] [\GeV]} \\
\hline
0.325(4) & 15.33(49) \times 10^{-12} & 4.98(12) \times 10^{-12} \\
\hline
1. & -0.732 & -0.480 \\
\nax & 1. & \phantom{-}0.949 \\
\nax & \nax & 1. \\
\hline\hline
\end{tabular}}
\caption{Central values, uncertainties and correlations for $R(\Lc)$, 
$\Gamma[\Lb \to \Lc \mu \nu]/|V_{cb}|^2$, and $\Gamma[\Lb \to \Lc \tau \nu]/|V_{cb}|^2$ in the \basescen fit scenario.}
\label{tab:RLcBFs}
\end{table}

\section{Summary}
\label{sec:summ}
In this work we developed the RCE description of $\Lb \to \Lc$ hadronic matrix elements.
We showed that in the $\Lb \to \Lc$ system, truncating the HQ expansion at $\mathcal{O}(\theta^2)$ in the RCE
leads to second-order power corrections that are described entirely by the hadron mass parameters plus (the quotient of) a single subsubleading Isgur-Wise function, $\hvphq$.
This result holds not just at $\mathcal{O}(1/m_{c}^2)$ but at $\mathcal{O}(1/m_{c,b}^2)$ in the HQ expansion.
Compared to the six subsubleading Isgur-Wise functions required in the full HQET description,
the RCE thus generates a highly-constrained and highly-predictive structure for the $\Lb \to \Lc$ form factors.
In particular, at zero recoil, the $\mathcal{O}(1/m_{c}^2)$ corrections are fully determined by the hadron mass parameters, $\LamB$ and $\lam1$,
which are in turn determined by the $\Lb$ and $\Lc$ baryon masses and the $1S$ scheme mass parameters (up to third-order corrections $\sim \rh1$).
The resulting RCE predictions for the SM form factors $g_{1}(1)$ and $f_{1}(1)$ at zero recoil 
and $\mathcal{O}(\aS^2, \aS/m_{c,b}, 1/m_{c,b}^2)$ are in excellent agreement with LQCD calculations.

We further showed that RCE-based form factor parametrizations including terms up to and including $\mathcal{O}(\aS^2, \aS/m_{c,b}, 1/m_{c,b}^2)$ 
yield fits in good agreement with LQCD predictions and LHCb data for the normalized $q^2$ spectrum,
and are compatible with prior BLRS parametrization fits, with comparable fit quality.
Thus even though the structure of the RCE-based parametrizations is more highly constrained than in BLRS, 
no biases appear to arise in their description of the matrix elements at second order.
The RCE-based $R(\Lc)$ predictions are similarly found to be compatible with prior BLRS results.
In particular, for our \basescen fit scenario we recover $R(\Lc) =  0.3251(37)$ in good agreement with the (modified) BLRS prediction $R(\Lc) = 0.3233(37)$~\cite{Bernlochner:2018kxh}.

Beyond the encouraging previous results for the RCE analysis of the $\Bbar \to D^{(*)}$ system~\cite{Bernlochner:2022ywh},
and especially given the sensitivity of the $\Lb \to \Lc$ system to second-order power corrections, 
the excellent performance of the RCE truncation scheme for the $\Lb \to \Lc$ system provides additional substantial evidence 
of its ability to faithfully capture (and provide a predictive framework for) the dominant second-order power corrections within HQET.

\acknowledgements
We thank Zoltan Ligeti for many insightful discussions and comments on the manuscript.
FB is supported by DFG Emmy-Noether Grant No.\ BE~6075/1-1 and BMBF Grant No.\ 05H21PDKBA. 
DJR is supported by the Office of High Energy Physics of the U.S.\ Department of Energy under contract DE-AC02-05CH11231. 
MP is supported by the U.S.\ Department of Energy, Office of High Energy Physics, under Award Number DE-SC0011632 and by the Walter Burke Institute for Theoretical Physics.

\appendix
\makeatletter
\let\save@section\section
\renewcommand{\section}[1]{
	\save@section{#1}
	\addtocontents{toc}{\protect\vspace*{-10pt}}
}
\makeatother

\section{$\Lb \to \Lc$ first and second-order power corrections}
\label{app:ffderivation}
Before proceeding to derive the second order corrections, it is useful to rederive the well-known first-order power corrections.
The first-order current corrections are determined by noting the HQET matrix elements must have the general form
\begin{subequations}
\label{eqn:currNLO}
\begin{align}
	\big\langle \Lc^{v'} \big| \cbvp \Gamma i \overrightarrow{D}_\mu \bv \big| \Lb^v \big\rangle & = \LamB\, \Ub^{v'} \Gamma\, \Us^v  \zeta_\mu(v,v')\,,\\*
	\big\langle \Lc^{v'} \big| \cbvp  (-i \overleftarrow{D}_\mu) \Gamma \bv \big| \Lb^v \big\rangle & = \LamB\, \Ub^{v'} \Gamma\, \Us^v \zeta^*_\mu(v',v)\,,
\end{align}
\end{subequations}
In the $s^{\pi_\ell}_\ell = 0^+$ HQET, $\zeta_\mu$ must be a scalar function,
with most general form $\zeta_\mu(v,v') = \zeta_+(v+v')_\mu + \zeta_-(v-v')_\mu$.
Because QCD and hence HQET is $PT$-invariant, 
the $PT$ symmetry of the $i\overrightarrow{D}_\mu$ operator ensures that $\zeta^*_\mu(v,v') = \zeta_\mu(v,v')$,
and therefore $\zeta_{\pm} = \zeta_{\pm}(w)$ must be real functions of $w$.
Straightforward application of the leading-order Schwinger-Dyson relation~\eqref{eqn:SDrelNLO} plus the equation of motion for $Q^v_+$---allows 
one to immediately deduce $\zeta_- = \zeta/2$, and $\zeta_+ = (w-1)/(w+1) \times \zeta/2$.

The first-order Lagrangian corrections are determined by writing the HQET matrix elements
\begin{subequations}
\label{eqn:chromoNLO}
\begin{align}
	\big\langle \Lc^{v'} \big| \cbvp \Gamma \bv \circ \big[\bbv D^2 \bv \big]\big| \Lb^v \big\rangle & = -\LamB\,  \Ub^{v'} \Gamma\, \Us^v X_0(v,v')\,,\\*
	\big\langle \Lc^{v'} \big| \cbvp \Gamma  \bv \circ \big[\bbv \frac{g}{2}\sigma_{\ab} G^{\ab} \bv \big] \big| \Lb^v \big\rangle & = -\LamB\, \Ub^{v'} \Gamma  \Pi_+ \sigma_{\ab}\, \Us^v X^{\ab}(v,v')\,,
\end{align}
\end{subequations}
where again $X_0$ and $X_{\ab}$ must be scalar functions 
(and noting the overall sign in Eq.~\eqref{eqn:mL1}).
The $PT$ symmetry of $D^2$ then ensures $X_0(v,v') = X^*_0(v,v')$,
and thus $X_0 \equiv 2\chi_1$ must be a real function of $w$.
Similarly, the $PT$ antisymmetry of $G^{\ab}$ ensures that $X_{\ab}(v,v') = -X^*_{\ab}(v, v')$.
However, as $X_{\ab}$ can have only the general form $\propto v_\alpha v'_\beta - v_\beta v'_\alpha$,
the identity $ \Pi_+ \sigma_{\ab}\Pi_+ v^\alpha = 0$ ensures that the chromomagnetic corrections all vanish.

Matching Eqs.~\eqref{eqn:currNLO} and~\eqref{eqn:chromoNLO} onto Eq.~\eqref{eqn:QCDmatchRC} one finds 
the first-order corrections
\begin{equation}
	\label{eqn:Ki1}
	\hK11 = 2\hat\chi_1\,,\qquad \hK21 = -1/(w+1)\,.
\end{equation}
These results are the same as in the standard HQ power expansion:
As can be seen in Eq.~\eqref{eqn:QCDmatchRC},
at $\mathcal{O}(\theta^2)$ in the RCE all first-order terms are retained.

The second-order current corrections from $\mathcal{J}_2$ arise from HQET matrix elements
\begin{subequations}
\label{eqn:currNNLO}
\begin{align}
	\big\langle \Lc^{v'} \big| \cbvp \Gamma D^2 \bv \big| \Lb^v \big\rangle & = -\LamB^2\, \Ub^{v'} \Gamma\, \Us^v \Phi_0(v,v')\,,\label{eqn:currNNLOD2}\\*
	\big\langle \Lc^{v'} \big| \cbvp \Gamma  \frac{g}{2}\sigma_{\ab} G^{\ab} \bv \big| \Lb^v \big\rangle & = -\LamB^2\,  \Ub^{v'} \Gamma \sigma_{\ab}\, \Us^v \Phi^{\ab}(v,v')]\,, \label{eqn:currNNLOG}
	\end{align}
\end{subequations}
and similarly for the conjugate matrix elements (and noting the overall sign in the definition of $\mJ_2$).
Once again, $\Phi_0$ and $\Phi_{\ab}$ must be scalar functions,
with most general forms $\Phi_0 = 2\varphi_0(w)$ and~\footnote{With respect to the notation of Ref.~\cite{Falk:1992ws}, 
$2\LamB^2\varphi_0 = \phi_0$ and $2\LamB^2 \varphi_1 = -\phi_1$.} 
\begin{equation}
	\Phi_{\ab}(v,v') = i\varphi_1(w) [v_\alpha v'_\beta - v_\beta v'_\alpha]\,.
\end{equation}
The $PT$ antisymmetry of $G^{\ab}$ requires that $\Phi_{\ab}(v,v') = -\Phi^*_{\ab}(v, v')$, 
so that $\varphi_1(w)$ is a real function of $w$. 
Similarly $PT$ symmetry of $D^2$ ensures $\Phi_0 = 2\varphi_0(w)$ is real.
Note there is no $\Pi_+$ projector in Eq.~\eqref{eqn:currNNLOG}, so that this matrix element does not vanish.
Matching Eqs~\eqref{eqn:currNNLO} onto Eq.~\eqref{eqn:QCDmatchRC}, one finds that 
\begin{equation}
	\label{eqn:K22}
	\hK22 = 2\hvph\,,
\end{equation}
while there is no current correction contribution to $\hK12$, as expected.

Although $\varphi_0$ does not explicitly appear in the second-order current corrections, 
the Schwinger-Dyson relations and heavy quark equation of motion generate interrelated constraints on $\varphi_0$ and $\varphi_1$.
First, applying the subleading Schwinger-Dyson relation~\eqref{eqn:SDrelNNLO} to Eqs~\eqref{eqn:currNNLO},
one finds that $\LamB^2 \Ub^{v'} \Gamma\, \Us^v 2\varphi_0 + \LamB^2 \Ub^{v'} \Gamma \Pi_+ \sigma_{\ab}\, \Us^v \Phi^{\ab}(v,v') = \lam1 \zeta \, \Ub^{v'} \Gamma\, \Us^v$.
The second term on the left hand side vanishes because of the $\Pi_+$ projector imposed by the Schwinger-Dyson relation, 
so that at $\mathcal{O}(\theta^2)$ in the RCE
\begin{equation}
	\varphi_0(w)  = \frac{\lam1}{2\LamB^2} \zeta(w)\,.
\end{equation}
(If instead all second-order power corrections are included, this relation holds only at zero recoil~\cite{Falk:1992ws}.)

To understand constraints on the Isgur-Wise function $\varphi_1$, 
one may employ an equivalent definition for $\Phi_{\ab}$, with respect to the matrix element 
\begin{equation}
	\label{eqn:currNNLODD}
	\big\langle \Lc^{v'} \big| \cbvp \Gamma  D_{\alpha} D_{\beta} \bv \big| \Lb^v \big\rangle 
		 = -\LamB^2\,\Ub^{v'} \Gamma\, \Us^v \big[\Psi_{\ab}(v,v') + i\Phi_{\ab}(v,v')\big]\,,
\end{equation}
in which the symmetric tensor $\Psi_{\ab}$ has the general form
\begin{equation}
	\Psi_{\ab}(v,v') = \psi_1g_{\ab} + \psi_2(v+v')_{\alpha}(v+v')_{\beta} + \psi_3(v-v')_{\alpha}(v-v')_{\beta} + \psi_4(v+v')_{(\alpha}(v-v')_{\beta)}\,,
\end{equation}
with $\psi_i$ real functions of $w$ by $PT$-invariance.
Comparing Eq.~\eqref{eqn:currNNLODD} with Eq.~\eqref{eqn:currNNLOD2}, it must be the case that $g^{\ab}\Psi_{\ab}(v,v') = 2\varphi_0$.
The heavy quark equation of motion acting on $\bv$, via contraction of Eq.~\eqref{eqn:currNNLODD} with $v^\beta$, requires that 
$v^\beta \big[\Psi_{\ab}(v,v') + i\Phi_{\ab}(v,v')\big] = 0$. 
Integration by parts, combined with the leading-order Schwinger-Dyson relation~\eqref{eqn:SDrelNLO},
requires further that
\begin{equation}
	\label{eqn:DDibp}
	\big\langle \Lc^{v'} \big| \cbvp (-i \overleftarrow{D}_\alpha) \Gamma  i D_{\beta} \bv \big| \Lb^v \big\rangle = 
	-\LamB\, (v-v')_\alpha \big\langle \Lc^{v'} \big| \cbvp \Gamma i D_{\beta} \bv \big| \Lb^v \big\rangle 
	-\big\langle \Lc^{v'} \big| \cbvp \Gamma  D_{\alpha} D_{\beta} \bv \big| \Lb^v \big\rangle\,.
\end{equation}
It then follows by the equation of motion for $\cbvp$, Eq.~\eqref{eqn:currNNLODD}, 
and Eq.~\eqref{eqn:currNLO} that $v^{\prime\alpha} \big[\Psi_{\ab}(v,v') + i\Phi_{\ab}(v,v')\big] = (w-1)\zeta_\beta(v,v')$.
These three conditions together have solution for the $\psi_i$'s,
\begin{align}
	\hat\psi_1 & = \frac{\lam1}{2\LamB^2}- w \hvph + \frac{w-1}{2 (w+1)}\,,\nn\\
	\hat\psi_2 & = - \frac{\lam1}{4 \LamB^2 (w+1)} + \frac{(2 w-1) \hvph}{2 (w+1)} + \frac{(w-1) (w-2)}{4 (w+1)^2}\,, \nn\\
	\hat\psi_3 & = \frac{\lam1}{4\LamB^2 (w-1)}  - \frac{(2 w+1) \hvph}{2(w-1)} + \frac{w+2}{4(w+1)}\,,\nn\\
	\hat\psi_4 & = \frac{w - 1}{4 (w+1)}\,. \label{eqn:psisolns}
\end{align} 
Similar to the argument presented in Appendix~D of Ref.~\cite{Bernlochner:2022ywh} (see also Refs.~\cite{Falk:1992ws,Falk:1992wt})
the product $(w-1)\psi_3(w)$ must vanish at zero recoil, so that $\hvph$ must obey the zero-recoil constraint
\begin{equation}
	\label{eqn:vphzerorec}
	\hvph(1) = \frac{\lam1}{6\LamB^2}\,.
\end{equation}
This zero-recoil constraint holds independently of the RCE, i.e. even when all second-order power corrections are included~\cite{Falk:1992ws}.
Analyticity of the matrix elements near zero recoil permits us to write
\begin{equation}
	\label{eqn:phpdef}
	\hvph(w) = \frac{\lam1}{6\LamB^2} + (w-1)\hvphq(w)\,.
\end{equation}
where the function $\hvphq$ must be regular. 
As in the definition in Eq.~\eqref{eqn:quotdef}, $\hvphq$ denotes the \emph{quotient} with respect to $w=1$.

The second-order product current corrections $\sim \mJbar'_1 \mJ_1$ follow directly from Eq.~\eqref{eqn:DDibp}, 
applying the solutions~\eqref{eqn:psisolns}. One finds
\begin{align}
	\label{eqn:Mi}
	\hM3 & = \frac{1}{w+1}\bigg[\frac{(w^2+2)\hvph }{(w-1)} - \frac{w \lam1 }{2\LamB^2(w-1)} - \frac{w-2}{2(w+1)} \bigg]\,,\nn\\
	\hM4 & = \frac{\lam1 }{2\LamB^2} - w\hvph + \frac{w-1}{2(w+1)}\,.
\end{align}
The zero-recoil constraint~\eqref{eqn:vphzerorec} ensures $\hM3$ is nonsingular at $w=1$. 

The second-order Lagrangian corrections from $\mathcal{L}_2$~\eqref{eqn:mL2} arise from HQET matrix elements 
\begin{subequations}
\label{eqn:chromoNNLO}
\begin{align}
	\big\langle \Lc^{v'} \big| \cbvp \Gamma \bv \circ \big[\bbv g v_\beta  D_\alpha G^{\ab}  \bv \big]\big| \Lb^v \big\rangle
		 & = \LamB^2\, \Ub^{v'} \Gamma\, \Us^v B_0(v,v')\,,\\*
	\big\langle \Lc^{v'} \big| \cbvp \Gamma  \bv \circ \big[-i\bbv v_\alpha \sigma_{\beta\gamma} D^\gamma G^{\ab} \bv \big] \big| \Lb^v \big\rangle 
		& = \LamB^2\, \Ub^{v'} \Gamma  \Pi_+ \sigma_{\ab}\, \Us^v B^{\ab}(v,v')\,.
\end{align}
\end{subequations}
Just as for the first-order corrections, the terms involving the tensor $B_{\ab}$ vanish, while $B_0 \equiv 2 \beta_1$ is a real function of $w$. 
Matching onto Eq.~\eqref{eqn:QCDmatchRC} yields simply
\begin{equation}
	\label{eqn:K12}
	\hK12 = 2\hat\beta_1\,.
\end{equation}
As derived in Sec.~\ref{sec:summff}, the mass normalization condition~\eqref{eqn:massnormcon} 
plus Eqs.~\eqref{eqn:K22}, \eqref{eqn:Mi}, \eqref{eqn:K12} 
and the zero-recoil constraint~\eqref{eqn:vphzerorec} imply that $\hat{\beta}_1(1) = \lam1/(4 \LamB^2)$.

\section{Radiative corrections}
\label{app:radcrr}
A general discussion of the $\mathcal{O}(\aS)$ and $\mathcal{O}(\aS/m_{c,b})$ corrections specific to the notation and conventions used in this work may be found in Ref.~\cite{Bernlochner:2022ywh},
following from long-known results~\cite{Neubert:1992qq,Neubert:1993iv,Neubert:1993mb}.

At $\mathcal{O}(\aS/m_{c,b})$, radiative corrections arise from local operators, 
which are in turn determined by reparametrization invariance, 
and nonlocal contributions, 
which are generated by operator products of $\mathcal{O}(\aS)$ perturbative corrections with insertions of the first-order Lagrangian $\mathcal{L}_1$. 
In the $\Lb \to \Lc$ system, the latter involve only kinetic energy terms proportional to $\chi_1$, 
and thus similarly to Eq.~\eqref{eqn:replNNLO} the nonlocal corrections may be reabsorbed into 
the leading-order Isgur-Wise function up to $\mathcal{O}(\aS/m^2_{c,b})$ corrections, that we do not consider. 

The local $\mathcal{O}(\aS/m_c)$ corrections introduce additional current contributions
\begin{multline}
	\delta \big[\cbar \Gamma b\big]
	 \to \cbvp \frac{\mJbar_1^{\prime}}{2m_c}\Big[ \haS \sum_i C_{\Gamma_i} \Gamma_i\Big] \bv \\*
	 +  \frac{1}{2m_c} \haS \sum_i \Big[2 \LamB (w-1)C'_{\Gamma_i}\,\cbvp \Gamma_i\bv + 2C_{\Gamma_i}\cbvp \big[\partial_{v'_\mu}\Gamma_i\big](-i \overleftarrow{D}_\mu)\bv\Big]\,.
\end{multline}
with $C'_\Gamma = \partial C_\Gamma /\partial w$.
The local $\mathcal{O}(\aS/m_b)$ corrections are constructed similarly via Hermitian conjugates of these terms. 
Incorporating these corrections via $\hat x_i \to \hat x_i + \haS \delta \hat x_i$ with respect to the form factor expressions in Eqs~\eqref{eqn:ffmatchwfn}
for base symbol $x = h$, $f$, or $g$, 
and applying the redefinition~\eqref{eqn:replNNLO},
one finds~\cite{Bernlochner:2018bfn,Bernlochner:2018kxh}
\begin{subequations}
\label{eqn:asmcorr}
\begin{align}
	\delta \hat h_S & =  (\ec + \eb) \big[ \Cs \, \frac{w-1}{w+1} + 2(w-1)\Cs'\big]\,,\\
	\delta \hat h_P & = (\ec + \eb) \big[ \Cps + 2(w-1)\Cps' \big]\,,\\
	\delta \hat f_1 & =  (\ec + \eb) \big[ \Cv1 + 2(w-1)\Cv1'\big]\,,\\
	\delta \hat f_2 & = \eb \Cv2 \frac{3w-1}{w+1} - \frac{\ec}{w+1}\big[2\Cv1 - (w-1) \Cv2 + 2\Cv3\big] + 2(w-1)(\ec + \eb)\Cv2' \,,\\
	\delta \hat f_3 & =  \ec \Cv3 \frac{3w-1}{w+1} - \frac{\eb}{w+1} \big[2\Cv1 + 2\Cv2 - (w-1) \Cv3\big] + 2(w-1)(\ec + \eb) \Cv3' \,\\
	\delta \hat g_1 & = (\ec + \eb) \Big[\Ca1\, \frac{w-1}{w+1} + 2(w-1)\Ca1'\Big]\,,\\
	\delta \hat g_2 & = \eb \Ca2 \frac{3w+1}{w+1}  - \frac{\ec}{w+1} \big[2\Ca1 - (w+1) \Ca2  + 2\Ca3\big] + 2(w-1)(\ec + \eb)\Ca2' \,,\\
	\delta \hat g_3 & =  \ec \Ca3 \frac{3w+1}{w+1} +  \frac{\eb}{w+1} \big[2\Ca1 - 2\Ca2 + (w+1) \Ca3\big] + 2(w-1)(\ec + \eb)\Ca3'\,,\\
	\delta \hat h_1 & = (\ec + \eb) \Big[ \Ct1 \frac{w-1}{w+1} + 2(w-1)\Ct1' \Big]\,,\\
	\delta \hat h_2 & = \eb \Ct2 \frac{3w+1}{w+1} - \frac{\ec}{w+1} \big[2\Ct1- (w+1) \Ct2 + 2\Ct3\big]+ 2(w-1)(\ec + \eb)\Ct2' \,,\\
	\delta \hat h_3 & = \ec \Ct3 \frac{3w+1}{w+1} +  \frac{\eb}{w+1}  \big[2\Ct1 - 2\Ct2 + (w+1) \Ct3\big] + 2(w-1)(\ec + \eb) \Ct3' \,,\\
	\delta \hat h_4 & = \frac{2}{w+1} \big[ \ec \Ct3  - \eb \Ct2 \big]\,.
\end{align}
\end{subequations}

\section{Zero-recoil NP predictions}
\label{app:NPzerorec}
At zero recoil and $\mathcal{O}(\aS, 1/m_{c,b}^2, \theta^2)$, the scalar, pseudoscalar and tensor current form factors have explicit form (noting $\hat{W}(1) = W(1)$ for each form factor)
\begin{subequations}
\begin{align}
h_S(1)_{\text{RC}}& = 1+\Cs \haS+ (\eb + \ec)^2  \lamL[][2]\,, \\
h_P(1)_{\text{RC}}& = 1+\Cps \haS + \eb+\ec - (\eb + \ec)^2  \lamL[][6]\,, \\
h_1(1)_{\text{RC}}& = 1+\Ct1 \haS+\eb^2 \lamL[][2]-\eb \ec \lamL[][3]+ \ec^2 \lamL[][2]\,, \\
h_2(1)_{\text{RC}}& = \Ct2 \haS -\ec -\eb \ec \bigg(\frac{1}{2}+6 \hvphp+\lamL[][3]\bigg)+ \ec^2\lamL[2][3]\,, \\
h_3(1)_{\text{RC}}& = \Ct3 \haS + \eb+\eb \ec \bigg(\frac{1}{2}+6 \hvphp+\lamL[][3]\bigg)- \eb^2\lamL[2][3]\,, \\*
h_4(1)_{\text{RC}}& = \eb \ec \bigg(\frac{1}{2}+6 \hvphp-\lamL[][3]\bigg)\,.
\end{align}
\end{subequations}
As for $g_1(1)$, the first-order corrections vanish for $h_1(1)$, so that its first-order prediction is precisely determined to be $h_1(1)_{\text{NLO}} = 1.030(1)$.
When including all second-order corrections analogously to Eq.~\eqref{eqn:FGexpnum}, 
the explicit numerical forms for the tensor form factors
\begin{subequations}
\begin{align}
h_1(1) & = 1.030 + 0.065\big[\lam1/\GeV^2\big]\,,\\*
h_2(1) & = -0.476 + (0.046 - 0.037 \hvphp)\big[\lam1/\GeV^2\big] - 0.199 \hvphp\,\\
h_3(1) & = 0.178 + (0.021 + 0.037\hvphp)\big[\lam1/\GeV^2\big] + 0.199 \hvphp\,,\\*
h_4(1) & = 0.036 + (0.037\hvphp-0.009)\big[\lam1/\GeV^2\big]  + 0.199 \hvphp\,.
\end{align}
\end{subequations}
As noted in Ref.~\cite{Bernlochner:2018kxh}, there is a substantial tension between the fit results for $\hat{h}_1(w)$ therein versus the LQCD calculation.
Similarly here, the LQCD prediction $h_1(1) = 0.86(4)$ 
substantially differs from the RCE-based second-order prediction over the nominal range $\lam1 = -0.1 \pm 0.2\,\GeV^2$:
One would require an implausibly large $\lam1$ and/or an implausibly large $\mathcal{O}(\aS^2)$ correction to account for this difference.

\section{Unremarkable correlations}
\label{sec:fitcorr}

\begin{table}[!h]
\newcolumntype{L}{ >{\arraybackslash $} l <{$}}
\newcolumntype{R}{ >{\raggedleft\arraybackslash $} m{1.5cm} <{$}}
\resizebox{0.8\textwidth}{!}{
\begin{tabular}{L|RRRRRRR}
\hline\hline
\text{\basescen} & \zeta' & \zeta'' & \mbS & \dmbc & \hvphp & \hvphpp & \rh1 \\ 
\hline
\zeta' & 1. & -0.784 & 0.247 & -0.041 & 0.145 & 0.122 & 0.026 \\ 
\zeta'' & \nax & 1. & -0.165 & 0.028 & -0.132 & 0.348 & -0.025 \\ 
\mbS & \nax & \nax & 1. & 0.097 & 0.031 & -0.146 & -0.013 \\ 
\dmbc & \nax & \nax & \nax & 1. & 0.037 & 0.014 & 0.119 \\ 
\hvphp & \nax & \nax & \nax & \nax & 1. & -0.507 & -0.042 \\ 
\hvphpp & \nax & \nax & \nax & \nax & \nax & 1. & -0.005 \\ 
\rh1 & \nax & \nax & \nax & \nax & \nax & \nax & 1. \\ 
\hline\hline
\end{tabular}}
\caption{Parameter correlations for the ``\basescen'' RCE-based fit scenario.}
\label{tab:fitcorrbasescen}
\end{table}

\begin{table}[!h]
\newcolumntype{L}{ >{\arraybackslash $} l <{$}}
\newcolumntype{R}{ >{\raggedleft\arraybackslash $} m{1.5cm} <{$}}
\resizebox{0.8\textwidth}{!}{
\begin{tabular}{L|RRRRRRR}
\hline\hline
\text{\basescen[\LOcubic]} & \zeta' & \zeta'' & \zeta''' & \mbS & \dmbc & \hvphp & \rh1 \\ 
\hline
\zeta' & 1. & -0.968 & 0.939 & 0.244 & 0.018 & -0.053 & -0.054 \\ 
\zeta'' & \nax & 1. & -0.993 & -0.187 & -0.034 & 0.154 & 0.067 \\ 
\zeta''' & \nax & \nax & 1. & 0.165 & 0.038 & -0.148 & -0.070 \\ 
\mbS & \nax & \nax & \nax & 1. & 0.098 & -0.049 & -0.031 \\ 
\dmbc & \nax & \nax & \nax & \nax & 1. & 0.051 & 0.120 \\ 
\hvphp & \nax & \nax & \nax & \nax & \nax & 1. & -0.054 \\ 
\rh1 & \nax & \nax & \nax & \nax & \nax & \nax & 1. \\ 
\hline\hline
\end{tabular}}
\caption{Parameter correlations for the ``\basescen[\LOcubic]'' RCE-based fit scenario.}
\label{tab:fitcorrbasescen[LOcubic]}
\end{table}

\begin{table}[!h]
\newcolumntype{L}{ >{\arraybackslash $} l <{$}}
\newcolumntype{R}{ >{\raggedleft\arraybackslash $} m{1.5cm} <{$}}
\resizebox{0.8\textwidth}{!}{
\begin{tabular}{L|RRRRRR}
\hline\hline
\text{\lqcdscen} & \zeta' & \zeta'' & \mbS & \dmbc & \hvphp & \rh1 \\ 
\hline
\zeta' & 1. & -0.805 & 0.171 & 0.055 & -0.014 & -0.086 \\ 
\zeta'' & \nax & 1. & -0.133 & -0.027 & 0.076 & 0.046 \\ 
\mbS & \nax & \nax & 1. & 0.086 & 0.004 & 0.001 \\ 
\dmbc & \nax & \nax & \nax & 1. & 0.023 & 0.097 \\ 
\hvphp & \nax & \nax & \nax & \nax & 1. & -0.021 \\ 
\rh1 & \nax & \nax & \nax & \nax & \nax & 1. \\ 
\hline\hline
\end{tabular}}
\caption{Parameter correlations for the ``\lqcdscen'' RCE-based fit scenario.}
\label{tab:fitcorrlqcdscen}
\end{table}

\begin{table}[!h]
\newcolumntype{L}{ >{\arraybackslash $} l <{$}}
\newcolumntype{R}{ >{\raggedleft\arraybackslash $} m{1.5cm} <{$}}
\resizebox{0.8\textwidth}{!}{
\begin{tabular}{L|RRRRRR}
\hline\hline
\text{BLRS} & \zeta' & \zeta'' & \mbS & \dmbc & \hat{b}_1 & \hat{b}_2 \\ 
\hline
\zeta' & 1. & -0.935 & 0.124 & -0.009 & -0.138 & 0.106 \\ 
\zeta'' & \nax & 1. & -0.117 & 0.004 & 0.119 & -0.029 \\ 
\mbS & \nax & \nax & 1. & -0.004 & -0.200 & -0.620 \\ 
\dmbc & \nax & \nax & \nax & 1. & 0.093 & 0.047 \\ 
\hat{b}_1 & \nax & \nax & \nax & \nax & 1. & 0.098 \\ 
\hat{b}_2 & \nax & \nax & \nax & \nax & \nax & 1. \\ 
\hline\hline
\end{tabular}}
\caption{Parameter correlations for the ``BLRS'' (modified) fit.}
\label{tab:fitcorrBLRS}
\end{table}

\begin{table}[!h]
\newcolumntype{L}{ >{\arraybackslash $} l <{$}}
\newcolumntype{R}{ >{\raggedleft\arraybackslash $} m{1.5cm} <{$}}
\resizebox{0.8\textwidth}{!}{
\begin{tabular}{L|RRRRRR}
\hline\hline
\text{\blrsscen[\LOcubic]} & \zeta' & \zeta'' & \zeta''' & \mbS & \dmbc & \hat{b}_1 \\ 
\hline
\zeta' & 1. & -0.976 & 0.949 & 0.215 & -0.011 & -0.354 \\ 
\zeta'' & \nax & 1. & -0.994 & -0.162 & 0.007 & 0.337 \\ 
\zeta''' & \nax & \nax & 1. & 0.142 & -0.006 & -0.326 \\ 
\mbS & \nax & \nax & \nax & 1. & 0.013 & -0.193 \\ 
\dmbc & \nax & \nax & \nax & \nax & 1. & 0.076 \\ 
\hat{b}_1 & \nax & \nax & \nax & \nax & \nax & 1. \\ 
\hline\hline
\end{tabular}}
\caption{Parameter correlations for the ``\blrsscen[\LOcubic]'' fit.}
\label{tab:fitcorrblrsscen[LOcubic]}
\end{table}


\begin{thebibliography}{33}%
\makeatletter
\providecommand \@ifxundefined [1]{%
 \@ifx{#1\undefined}
}%
\providecommand \@ifnum [1]{%
 \ifnum #1\expandafter \@firstoftwo
 \else \expandafter \@secondoftwo
 \fi
}%
\providecommand \@ifx [1]{%
 \ifx #1\expandafter \@firstoftwo
 \else \expandafter \@secondoftwo
 \fi
}%
\providecommand \natexlab [1]{#1}%
\providecommand \enquote  [1]{``#1''}%
\providecommand \bibnamefont  [1]{#1}%
\providecommand \bibfnamefont [1]{#1}%
\providecommand \citenamefont [1]{#1}%
\providecommand \href@noop [0]{\@secondoftwo}%
\providecommand \href [0]{\begingroup \@sanitize@url \@href}%
\providecommand \@href[1]{\@@startlink{#1}\@@href}%
\providecommand \@@href[1]{\endgroup#1\@@endlink}%
\providecommand \@sanitize@url [0]{\catcode `\\12\catcode `\$12\catcode
  `\&12\catcode `\#12\catcode `\^12\catcode `\_12\catcode `\%12\relax}%
\providecommand \@@startlink[1]{}%
\providecommand \@@endlink[0]{}%
\providecommand \url  [0]{\begingroup\@sanitize@url \@url }%
\providecommand \@url [1]{\endgroup\@href {#1}{\urlprefix }}%
\providecommand \urlprefix  [0]{URL }%
\providecommand \Eprint [0]{\href }%
\providecommand \doibase [0]{https://doi.org/}%
\providecommand \selectlanguage [0]{\@gobble}%
\providecommand \bibinfo  [0]{\@secondoftwo}%
\providecommand \bibfield  [0]{\@secondoftwo}%
\providecommand \translation [1]{[#1]}%
\providecommand \BibitemOpen [0]{}%
\providecommand \bibitemStop [0]{}%
\providecommand \bibitemNoStop [0]{.\EOS\space}%
\providecommand \EOS [0]{\spacefactor3000\relax}%
\providecommand \BibitemShut  [1]{\csname bibitem#1\endcsname}%
\let\auto@bib@innerbib\@empty
\bibitem [{\citenamefont {Georgi}\ \emph {et~al.}(1990)\citenamefont {Georgi},
  \citenamefont {Grinstein},\ and\ \citenamefont {Wise}}]{Georgi:1990ei}%
  \BibitemOpen
  \bibfield  {author} {\bibinfo {author} {\bibfnamefont {H.}~\bibnamefont
  {Georgi}}, \bibinfo {author} {\bibfnamefont {B.}~\bibnamefont {Grinstein}},\
  and\ \bibinfo {author} {\bibfnamefont {M.~B.}\ \bibnamefont {Wise}},\
  }\bibfield  {title} {\bibinfo {title} {{Lambda(b) semileptonic decay
  form-factors for m(c) does not equal infinity}},\ }\href
  {https://doi.org/10.1016/0370-2693(90)90569-R} {\bibfield  {journal}
  {\bibinfo  {journal} {Phys. Lett. B}\ }\textbf {\bibinfo {volume} {252}},\
  \bibinfo {pages} {456} (\bibinfo {year} {1990})}\BibitemShut {NoStop}%
\bibitem [{\citenamefont {Falk}\ and\ \citenamefont
  {Neubert}(1993{\natexlab{a}})}]{Falk:1992ws}%
  \BibitemOpen
  \bibfield  {author} {\bibinfo {author} {\bibfnamefont {A.~F.}\ \bibnamefont
  {Falk}}\ and\ \bibinfo {author} {\bibfnamefont {M.}~\bibnamefont {Neubert}},\
  }\bibfield  {title} {\bibinfo {title} {{Second order power corrections in the
  heavy quark effective theory. 2. Baryon form-factors}},\ }\href
  {https://doi.org/10.1103/PhysRevD.47.2982} {\bibfield  {journal} {\bibinfo
  {journal} {Phys. Rev. D}\ }\textbf {\bibinfo {volume} {47}},\ \bibinfo
  {pages} {2982} (\bibinfo {year} {1993}{\natexlab{a}})},\ \Eprint
  {https://arxiv.org/abs/hep-ph/9209269} {arXiv:hep-ph/9209269} \BibitemShut
  {NoStop}%
\bibitem [{\citenamefont {Bernlochner}\ \emph {et~al.}(2018)\citenamefont
  {Bernlochner}, \citenamefont {Ligeti}, \citenamefont {Robinson},\ and\
  \citenamefont {Sutcliffe}}]{Bernlochner:2018kxh}%
  \BibitemOpen
  \bibfield  {author} {\bibinfo {author} {\bibfnamefont {F.~U.}\ \bibnamefont
  {Bernlochner}}, \bibinfo {author} {\bibfnamefont {Z.}~\bibnamefont {Ligeti}},
  \bibinfo {author} {\bibfnamefont {D.~J.}\ \bibnamefont {Robinson}},\ and\
  \bibinfo {author} {\bibfnamefont {W.~L.}\ \bibnamefont {Sutcliffe}},\
  }\bibfield  {title} {\bibinfo {title} {{New predictions for
  $\Lambda_b\to\Lambda_c$ semileptonic decays and tests of heavy quark
  symmetry}},\ }\href {https://doi.org/10.1103/PhysRevLett.121.202001}
  {\bibfield  {journal} {\bibinfo  {journal} {Phys. Rev. Lett.}\ }\textbf
  {\bibinfo {volume} {121}},\ \bibinfo {pages} {202001} (\bibinfo {year}
  {2018})},\ \Eprint {https://arxiv.org/abs/1808.09464} {arXiv:1808.09464
  [hep-ph]} \BibitemShut {NoStop}%
\bibitem [{\citenamefont {Bernlochner}\ \emph {et~al.}(2019)\citenamefont
  {Bernlochner}, \citenamefont {Ligeti}, \citenamefont {Robinson},\ and\
  \citenamefont {Sutcliffe}}]{Bernlochner:2018bfn}%
  \BibitemOpen
  \bibfield  {author} {\bibinfo {author} {\bibfnamefont {F.~U.}\ \bibnamefont
  {Bernlochner}}, \bibinfo {author} {\bibfnamefont {Z.}~\bibnamefont {Ligeti}},
  \bibinfo {author} {\bibfnamefont {D.~J.}\ \bibnamefont {Robinson}},\ and\
  \bibinfo {author} {\bibfnamefont {W.~L.}\ \bibnamefont {Sutcliffe}},\
  }\bibfield  {title} {\bibinfo {title} {{Precise predictions for $\Lambda_b
  \to \Lambda_c$ semileptonic decays}},\ }\href
  {https://doi.org/10.1103/PhysRevD.99.055008} {\bibfield  {journal} {\bibinfo
  {journal} {Phys. Rev. D}\ }\textbf {\bibinfo {volume} {99}},\ \bibinfo
  {pages} {055008} (\bibinfo {year} {2019})},\ \Eprint
  {https://arxiv.org/abs/1812.07593} {arXiv:1812.07593 [hep-ph]} \BibitemShut
  {NoStop}%
\bibitem [{\citenamefont {Detmold}\ \emph {et~al.}(2015)\citenamefont
  {Detmold}, \citenamefont {Lehner},\ and\ \citenamefont
  {Meinel}}]{Detmold:2015aaa}%
  \BibitemOpen
  \bibfield  {author} {\bibinfo {author} {\bibfnamefont {W.}~\bibnamefont
  {Detmold}}, \bibinfo {author} {\bibfnamefont {C.}~\bibnamefont {Lehner}},\
  and\ \bibinfo {author} {\bibfnamefont {S.}~\bibnamefont {Meinel}},\
  }\bibfield  {title} {\bibinfo {title} {{$\Lambda_b \to p \ell^-
  \bar{\nu}_\ell$ and $\Lambda_b \to \Lambda_c \ell^- \bar{\nu}_\ell$ form
  factors from lattice QCD with relativistic heavy quarks}},\ }\href
  {https://doi.org/10.1103/PhysRevD.92.034503} {\bibfield  {journal} {\bibinfo
  {journal} {Phys. Rev.}\ }\textbf {\bibinfo {volume} {D92}},\ \bibinfo {pages}
  {034503} (\bibinfo {year} {2015})},\ \Eprint
  {https://arxiv.org/abs/1503.01421} {arXiv:1503.01421 [hep-lat]} \BibitemShut
  {NoStop}%
\bibitem [{\citenamefont {Aaij}\ \emph {et~al.}(2017)\citenamefont {Aaij} \emph
  {et~al.}}]{Aaij:2017svr}%
  \BibitemOpen
  \bibfield  {author} {\bibinfo {author} {\bibfnamefont {R.}~\bibnamefont
  {Aaij}} \emph {et~al.} (\bibinfo {collaboration} {LHCb Collaboration}),\
  }\bibfield  {title} {\bibinfo {title} {{Measurement of the shape of the
  $\Lambda_b^0\to\Lambda_c^+ \mu^- \overline{\nu}_{\mu}$ differential decay
  rate}},\ }\href {https://doi.org/10.1103/PhysRevD.96.112005} {\bibfield
  {journal} {\bibinfo  {journal} {Phys. Rev.}\ }\textbf {\bibinfo {volume}
  {D96}},\ \bibinfo {pages} {112005} (\bibinfo {year} {2017})},\ \Eprint
  {https://arxiv.org/abs/1709.01920} {arXiv:1709.01920 [hep-ex]} \BibitemShut
  {NoStop}%
\bibitem [{\citenamefont {Bernlochner}\ \emph {et~al.}(2022)\citenamefont
  {Bernlochner}, \citenamefont {Ligeti}, \citenamefont {Papucci}, \citenamefont
  {Prim}, \citenamefont {Robinson},\ and\ \citenamefont
  {Xiong}}]{Bernlochner:2022ywh}%
  \BibitemOpen
  \bibfield  {author} {\bibinfo {author} {\bibfnamefont {F.~U.}\ \bibnamefont
  {Bernlochner}}, \bibinfo {author} {\bibfnamefont {Z.}~\bibnamefont {Ligeti}},
  \bibinfo {author} {\bibfnamefont {M.}~\bibnamefont {Papucci}}, \bibinfo
  {author} {\bibfnamefont {M.~T.}\ \bibnamefont {Prim}}, \bibinfo {author}
  {\bibfnamefont {D.~J.}\ \bibnamefont {Robinson}},\ and\ \bibinfo {author}
  {\bibfnamefont {C.}~\bibnamefont {Xiong}},\ }\bibfield  {title} {\bibinfo
  {title} {{Constrained second-order power corrections in HQET: $R(D^{(*)})$,
  $|V_{cb}|$, and new physics}},\ }\href
  {https://doi.org/10.1103/PhysRevD.106.096015} {\bibfield  {journal} {\bibinfo
   {journal} {Phys. Rev. D}\ }\textbf {\bibinfo {volume} {106}},\ \bibinfo
  {pages} {096015} (\bibinfo {year} {2022})},\ \Eprint
  {https://arxiv.org/abs/2206.11281} {arXiv:2206.11281 [hep-ph]} \BibitemShut
  {NoStop}%
\bibitem [{\citenamefont {Di~Risi}\ \emph {et~al.}(2023)\citenamefont
  {Di~Risi}, \citenamefont {Iacobacci},\ and\ \citenamefont
  {Sannino}}]{DiRisi:2023npw}%
  \BibitemOpen
  \bibfield  {author} {\bibinfo {author} {\bibfnamefont {V.}~\bibnamefont
  {Di~Risi}}, \bibinfo {author} {\bibfnamefont {D.}~\bibnamefont {Iacobacci}},\
  and\ \bibinfo {author} {\bibfnamefont {F.}~\bibnamefont {Sannino}},\
  }\bibfield  {title} {\bibinfo {title} {{$\Lambda_b \rightarrow
  \Lambda_c^{\ast}$ at $1\,/\,m_c^2$ heavy quark order}},\ }\href@noop {} {\
  (\bibinfo {year} {2023})},\ \Eprint {https://arxiv.org/abs/2309.03553}
  {arXiv:2309.03553 [hep-ph]} \BibitemShut {NoStop}%
\bibitem [{\citenamefont {Isgur}\ and\ \citenamefont
  {Wise}(1991)}]{Isgur:1990pm}%
  \BibitemOpen
  \bibfield  {author} {\bibinfo {author} {\bibfnamefont {N.}~\bibnamefont
  {Isgur}}\ and\ \bibinfo {author} {\bibfnamefont {M.~B.}\ \bibnamefont
  {Wise}},\ }\bibfield  {title} {\bibinfo {title} {{Heavy baryon weak
  form-factors}},\ }\href {https://doi.org/10.1016/0550-3213(91)90518-3}
  {\bibfield  {journal} {\bibinfo  {journal} {Nucl. Phys.}\ }\textbf {\bibinfo
  {volume} {B348}},\ \bibinfo {pages} {276} (\bibinfo {year}
  {1991})}\BibitemShut {NoStop}%
\bibitem [{\citenamefont {Manohar}\ and\ \citenamefont
  {Wise}(2000)}]{Manohar:2000dt}%
  \BibitemOpen
  \bibfield  {author} {\bibinfo {author} {\bibfnamefont {A.~V.}\ \bibnamefont
  {Manohar}}\ and\ \bibinfo {author} {\bibfnamefont {M.~B.}\ \bibnamefont
  {Wise}},\ }\href {https://doi.org/10.1017/CBO9780511529351} {\emph {\bibinfo
  {title} {Heavy Quark Physics}}},\ Camb.\ Monogr.\ on Part.\ Phys., Nucl.\
  Phys., Cosmol.\ (\bibinfo  {publisher} {Cambridge University Press},\
  \bibinfo {year} {2000})\BibitemShut {NoStop}%
\bibitem [{\citenamefont {Neubert}(1992)}]{Neubert:1992qq}%
  \BibitemOpen
  \bibfield  {author} {\bibinfo {author} {\bibfnamefont {M.}~\bibnamefont
  {Neubert}},\ }\bibfield  {title} {\bibinfo {title} {{Renormalization of heavy
  quark currents}},\ }\href {https://doi.org/10.1016/0550-3213(92)90233-2}
  {\bibfield  {journal} {\bibinfo  {journal} {Nucl. Phys.}\ }\textbf {\bibinfo
  {volume} {B371}},\ \bibinfo {pages} {149} (\bibinfo {year}
  {1992})}\BibitemShut {NoStop}%
\bibitem [{\citenamefont {Bernlochner}\ \emph {et~al.}(2017)\citenamefont
  {Bernlochner}, \citenamefont {Ligeti}, \citenamefont {Papucci},\ and\
  \citenamefont {Robinson}}]{Bernlochner:2017jka}%
  \BibitemOpen
  \bibfield  {author} {\bibinfo {author} {\bibfnamefont {F.~U.}\ \bibnamefont
  {Bernlochner}}, \bibinfo {author} {\bibfnamefont {Z.}~\bibnamefont {Ligeti}},
  \bibinfo {author} {\bibfnamefont {M.}~\bibnamefont {Papucci}},\ and\ \bibinfo
  {author} {\bibfnamefont {D.~J.}\ \bibnamefont {Robinson}},\ }\bibfield
  {title} {\bibinfo {title} {{Combined analysis of semileptonic $B$ decays to
  $D$ and $D^*$: $R(D^{(*)})$, $|V_{cb}|$, and new physics}},\ }\href
  {https://doi.org/10.1103/PhysRevD.95.115008} {\bibfield  {journal} {\bibinfo
  {journal} {Phys. Rev.}\ }\textbf {\bibinfo {volume} {D95}},\ \bibinfo {pages}
  {115008} (\bibinfo {year} {2017})},\ \Eprint
  {https://arxiv.org/abs/1703.05330} {arXiv:1703.05330 [hep-ph]} \BibitemShut
  {NoStop}%
\bibitem [{\citenamefont {Hoang}\ \emph
  {et~al.}(1999{\natexlab{a}})\citenamefont {Hoang}, \citenamefont {Ligeti},\
  and\ \citenamefont {Manohar}}]{Hoang:1998ng}%
  \BibitemOpen
  \bibfield  {author} {\bibinfo {author} {\bibfnamefont {A.~H.}\ \bibnamefont
  {Hoang}}, \bibinfo {author} {\bibfnamefont {Z.}~\bibnamefont {Ligeti}},\ and\
  \bibinfo {author} {\bibfnamefont {A.~V.}\ \bibnamefont {Manohar}},\
  }\bibfield  {title} {\bibinfo {title} {{B decay and the Upsilon mass}},\
  }\href {https://doi.org/10.1103/PhysRevLett.82.277} {\bibfield  {journal}
  {\bibinfo  {journal} {Phys. Rev. Lett.}\ }\textbf {\bibinfo {volume} {82}},\
  \bibinfo {pages} {277} (\bibinfo {year} {1999}{\natexlab{a}})},\ \Eprint
  {https://arxiv.org/abs/hep-ph/9809423} {arXiv:hep-ph/9809423} \BibitemShut
  {NoStop}%
\bibitem [{\citenamefont {Hoang}\ \emph
  {et~al.}(1999{\natexlab{b}})\citenamefont {Hoang}, \citenamefont {Ligeti},\
  and\ \citenamefont {Manohar}}]{Hoang:1998hm}%
  \BibitemOpen
  \bibfield  {author} {\bibinfo {author} {\bibfnamefont {A.~H.}\ \bibnamefont
  {Hoang}}, \bibinfo {author} {\bibfnamefont {Z.}~\bibnamefont {Ligeti}},\ and\
  \bibinfo {author} {\bibfnamefont {A.~V.}\ \bibnamefont {Manohar}},\
  }\bibfield  {title} {\bibinfo {title} {{B decays in the upsilon expansion}},\
  }\href {https://doi.org/10.1103/PhysRevD.59.074017} {\bibfield  {journal}
  {\bibinfo  {journal} {Phys. Rev.}\ }\textbf {\bibinfo {volume} {D59}},\
  \bibinfo {pages} {074017} (\bibinfo {year} {1999}{\natexlab{b}})},\ \Eprint
  {https://arxiv.org/abs/hep-ph/9811239} {arXiv:hep-ph/9811239} \BibitemShut
  {NoStop}%
\bibitem [{\citenamefont {Hoang}(2000)}]{Hoang:1999ye}%
  \BibitemOpen
  \bibfield  {author} {\bibinfo {author} {\bibfnamefont {A.~H.}\ \bibnamefont
  {Hoang}},\ }\bibfield  {title} {\bibinfo {title} {{1S and MS-bar bottom quark
  masses from Upsilon sum rules}},\ }\href
  {https://doi.org/10.1103/PhysRevD.61.034005} {\bibfield  {journal} {\bibinfo
  {journal} {Phys. Rev.}\ }\textbf {\bibinfo {volume} {D61}},\ \bibinfo {pages}
  {034005} (\bibinfo {year} {2000})},\ \Eprint
  {https://arxiv.org/abs/hep-ph/9905550} {arXiv:hep-ph/9905550} \BibitemShut
  {NoStop}%
\bibitem [{\citenamefont {Bigi}\ \emph {et~al.}(1994)\citenamefont {Bigi},
  \citenamefont {Shifman}, \citenamefont {Uraltsev},\ and\ \citenamefont
  {Vainshtein}}]{Bigi:1994em}%
  \BibitemOpen
  \bibfield  {author} {\bibinfo {author} {\bibfnamefont {I.~I.~Y.}\
  \bibnamefont {Bigi}}, \bibinfo {author} {\bibfnamefont {M.~A.}\ \bibnamefont
  {Shifman}}, \bibinfo {author} {\bibfnamefont {N.~G.}\ \bibnamefont
  {Uraltsev}},\ and\ \bibinfo {author} {\bibfnamefont {A.~I.}\ \bibnamefont
  {Vainshtein}},\ }\bibfield  {title} {\bibinfo {title} {{The Pole mass of the
  heavy quark. Perturbation theory and beyond}},\ }\href
  {https://doi.org/10.1103/PhysRevD.50.2234} {\bibfield  {journal} {\bibinfo
  {journal} {Phys. Rev. D}\ }\textbf {\bibinfo {volume} {50}},\ \bibinfo
  {pages} {2234} (\bibinfo {year} {1994})},\ \Eprint
  {https://arxiv.org/abs/hep-ph/9402360} {arXiv:hep-ph/9402360} \BibitemShut
  {NoStop}%
\bibitem [{\citenamefont {Beneke}\ and\ \citenamefont
  {Braun}(1994)}]{Beneke:1994sw}%
  \BibitemOpen
  \bibfield  {author} {\bibinfo {author} {\bibfnamefont {M.}~\bibnamefont
  {Beneke}}\ and\ \bibinfo {author} {\bibfnamefont {V.~M.}\ \bibnamefont
  {Braun}},\ }\bibfield  {title} {\bibinfo {title} {{Heavy quark effective
  theory beyond perturbation theory: Renormalons, the pole mass and the
  residual mass term}},\ }\href {https://doi.org/10.1016/0550-3213(94)90314-X}
  {\bibfield  {journal} {\bibinfo  {journal} {Nucl. Phys. B}\ }\textbf
  {\bibinfo {volume} {426}},\ \bibinfo {pages} {301} (\bibinfo {year}
  {1994})},\ \Eprint {https://arxiv.org/abs/hep-ph/9402364}
  {arXiv:hep-ph/9402364} \BibitemShut {NoStop}%
\bibitem [{\citenamefont {Beneke}\ \emph {et~al.}(1994)\citenamefont {Beneke},
  \citenamefont {Braun},\ and\ \citenamefont {Zakharov}}]{Beneke:1994bc}%
  \BibitemOpen
  \bibfield  {author} {\bibinfo {author} {\bibfnamefont {M.}~\bibnamefont
  {Beneke}}, \bibinfo {author} {\bibfnamefont {V.~M.}\ \bibnamefont {Braun}},\
  and\ \bibinfo {author} {\bibfnamefont {V.~I.}\ \bibnamefont {Zakharov}},\
  }\bibfield  {title} {\bibinfo {title} {{Bloch-Nordsieck cancellations beyond
  logarithms in heavy particle decays}},\ }\href
  {https://doi.org/10.1103/PhysRevLett.73.3058} {\bibfield  {journal} {\bibinfo
   {journal} {Phys. Rev. Lett.}\ }\textbf {\bibinfo {volume} {73}},\ \bibinfo
  {pages} {3058} (\bibinfo {year} {1994})},\ \Eprint
  {https://arxiv.org/abs/hep-ph/9405304} {arXiv:hep-ph/9405304} \BibitemShut
  {NoStop}%
\bibitem [{\citenamefont {Neubert}\ and\ \citenamefont
  {Sachrajda}(1995)}]{Neubert:1994wq}%
  \BibitemOpen
  \bibfield  {author} {\bibinfo {author} {\bibfnamefont {M.}~\bibnamefont
  {Neubert}}\ and\ \bibinfo {author} {\bibfnamefont {C.~T.}\ \bibnamefont
  {Sachrajda}},\ }\bibfield  {title} {\bibinfo {title} {{Cancellation of
  renormalon ambiguities in the heavy quark effective theory}},\ }\href
  {https://doi.org/10.1016/0550-3213(95)00032-N} {\bibfield  {journal}
  {\bibinfo  {journal} {Nucl. Phys. B}\ }\textbf {\bibinfo {volume} {438}},\
  \bibinfo {pages} {235} (\bibinfo {year} {1995})},\ \Eprint
  {https://arxiv.org/abs/hep-ph/9407394} {arXiv:hep-ph/9407394} \BibitemShut
  {NoStop}%
\bibitem [{\citenamefont {Luke}\ \emph {et~al.}(1995)\citenamefont {Luke},
  \citenamefont {Manohar},\ and\ \citenamefont {Savage}}]{Luke:1994xd}%
  \BibitemOpen
  \bibfield  {author} {\bibinfo {author} {\bibfnamefont {M.~E.}\ \bibnamefont
  {Luke}}, \bibinfo {author} {\bibfnamefont {A.~V.}\ \bibnamefont {Manohar}},\
  and\ \bibinfo {author} {\bibfnamefont {M.~J.}\ \bibnamefont {Savage}},\
  }\bibfield  {title} {\bibinfo {title} {{Renormalons in effective field
  theories}},\ }\href {https://doi.org/10.1103/PhysRevD.51.4924} {\bibfield
  {journal} {\bibinfo  {journal} {Phys. Rev. D}\ }\textbf {\bibinfo {volume}
  {51}},\ \bibinfo {pages} {4924} (\bibinfo {year} {1995})},\ \Eprint
  {https://arxiv.org/abs/hep-ph/9407407} {arXiv:hep-ph/9407407} \BibitemShut
  {NoStop}%
\bibitem [{\citenamefont {Czakon}(2005)}]{Czakon:2004bu}%
  \BibitemOpen
  \bibfield  {author} {\bibinfo {author} {\bibfnamefont {M.}~\bibnamefont
  {Czakon}},\ }\bibfield  {title} {\bibinfo {title} {{The Four-loop QCD
  beta-function and anomalous dimensions}},\ }\href
  {https://doi.org/10.1016/j.nuclphysb.2005.01.012} {\bibfield  {journal}
  {\bibinfo  {journal} {Nucl. Phys. B}\ }\textbf {\bibinfo {volume} {710}},\
  \bibinfo {pages} {485} (\bibinfo {year} {2005})},\ \Eprint
  {https://arxiv.org/abs/hep-ph/0411261} {arXiv:hep-ph/0411261} \BibitemShut
  {NoStop}%
\bibitem [{\citenamefont {{van Ritbergen}}\ \emph {et~al.}(1997)\citenamefont
  {{van Ritbergen}}, \citenamefont {Vermaseren},\ and\ \citenamefont
  {Larin}}]{VANRITBERGEN1997379}%
  \BibitemOpen
  \bibfield  {author} {\bibinfo {author} {\bibfnamefont {T.}~\bibnamefont {{van
  Ritbergen}}}, \bibinfo {author} {\bibfnamefont {J.}~\bibnamefont
  {Vermaseren}},\ and\ \bibinfo {author} {\bibfnamefont {S.}~\bibnamefont
  {Larin}},\ }\bibfield  {title} {\bibinfo {title} {The four-loop
  $\beta$-function in quantum chromodynamics},\ }\href
  {https://doi.org/https://doi.org/10.1016/S0370-2693(97)00370-5} {\bibfield
  {journal} {\bibinfo  {journal} {Phys. Lett. B}\ }\textbf {\bibinfo {volume}
  {400}},\ \bibinfo {pages} {379} (\bibinfo {year} {1997})}\BibitemShut
  {NoStop}%
\bibitem [{\citenamefont {Bauer}\ \emph {et~al.}(2003)\citenamefont {Bauer},
  \citenamefont {Ligeti}, \citenamefont {Luke},\ and\ \citenamefont
  {Manohar}}]{Bauer:2002sh}%
  \BibitemOpen
  \bibfield  {author} {\bibinfo {author} {\bibfnamefont {C.~W.}\ \bibnamefont
  {Bauer}}, \bibinfo {author} {\bibfnamefont {Z.}~\bibnamefont {Ligeti}},
  \bibinfo {author} {\bibfnamefont {M.}~\bibnamefont {Luke}},\ and\ \bibinfo
  {author} {\bibfnamefont {A.~V.}\ \bibnamefont {Manohar}},\ }\bibfield
  {title} {\bibinfo {title} {{B decay shape variables and the precision
  determination of |V(cb)| and m(b)}},\ }\href
  {https://doi.org/10.1103/PhysRevD.67.054012} {\bibfield  {journal} {\bibinfo
  {journal} {Phys. Rev. D}\ }\textbf {\bibinfo {volume} {67}},\ \bibinfo
  {pages} {054012} (\bibinfo {year} {2003})},\ \Eprint
  {https://arxiv.org/abs/hep-ph/0210027} {arXiv:hep-ph/0210027} \BibitemShut
  {NoStop}%
\bibitem [{\citenamefont {Bauer}\ \emph {et~al.}(2004)\citenamefont {Bauer},
  \citenamefont {Ligeti}, \citenamefont {Luke}, \citenamefont {Manohar},\ and\
  \citenamefont {Trott}}]{Bauer:2004ve}%
  \BibitemOpen
  \bibfield  {author} {\bibinfo {author} {\bibfnamefont {C.~W.}\ \bibnamefont
  {Bauer}}, \bibinfo {author} {\bibfnamefont {Z.}~\bibnamefont {Ligeti}},
  \bibinfo {author} {\bibfnamefont {M.}~\bibnamefont {Luke}}, \bibinfo {author}
  {\bibfnamefont {A.~V.}\ \bibnamefont {Manohar}},\ and\ \bibinfo {author}
  {\bibfnamefont {M.}~\bibnamefont {Trott}},\ }\bibfield  {title} {\bibinfo
  {title} {{Global analysis of inclusive B decays}},\ }\href
  {https://doi.org/10.1103/PhysRevD.70.094017} {\bibfield  {journal} {\bibinfo
  {journal} {Phys. Rev. D}\ }\textbf {\bibinfo {volume} {70}},\ \bibinfo
  {pages} {094017} (\bibinfo {year} {2004})},\ \Eprint
  {https://arxiv.org/abs/hep-ph/0408002} {arXiv:hep-ph/0408002} \BibitemShut
  {NoStop}%
\bibitem [{\citenamefont {Ligeti}\ and\ \citenamefont
  {Tackmann}(2014)}]{Ligeti:2014kia}%
  \BibitemOpen
  \bibfield  {author} {\bibinfo {author} {\bibfnamefont {Z.}~\bibnamefont
  {Ligeti}}\ and\ \bibinfo {author} {\bibfnamefont {F.~J.}\ \bibnamefont
  {Tackmann}},\ }\bibfield  {title} {\bibinfo {title} {{Precise predictions for
  $B \to X_c \tau \bar \nu$ decay distributions}},\ }\href
  {https://doi.org/10.1103/PhysRevD.90.034021} {\bibfield  {journal} {\bibinfo
  {journal} {Phys. Rev.}\ }\textbf {\bibinfo {volume} {D90}},\ \bibinfo {pages}
  {034021} (\bibinfo {year} {2014})},\ \Eprint
  {https://arxiv.org/abs/1406.7013} {arXiv:1406.7013 [hep-ph]} \BibitemShut
  {NoStop}%
\bibitem [{\citenamefont {Gremm}\ and\ \citenamefont
  {Kapustin}(1997)}]{Gremm:1996df}%
  \BibitemOpen
  \bibfield  {author} {\bibinfo {author} {\bibfnamefont {M.}~\bibnamefont
  {Gremm}}\ and\ \bibinfo {author} {\bibfnamefont {A.}~\bibnamefont
  {Kapustin}},\ }\bibfield  {title} {\bibinfo {title} {{Order 1/m(b)**3
  corrections to B --\ensuremath{>} X(c) lepton anti-neutrino decay and their
  implication for the measurement of Lambda-bar and lambda(1)}},\ }\href
  {https://doi.org/10.1103/PhysRevD.55.6924} {\bibfield  {journal} {\bibinfo
  {journal} {Phys. Rev. D}\ }\textbf {\bibinfo {volume} {55}},\ \bibinfo
  {pages} {6924} (\bibinfo {year} {1997})},\ \Eprint
  {https://arxiv.org/abs/hep-ph/9603448} {arXiv:hep-ph/9603448} \BibitemShut
  {NoStop}%
\bibitem [{\citenamefont {Czarnecki}(1996)}]{Czarnecki:1996gu}%
  \BibitemOpen
  \bibfield  {author} {\bibinfo {author} {\bibfnamefont {A.}~\bibnamefont
  {Czarnecki}},\ }\bibfield  {title} {\bibinfo {title} {{Two loop QCD
  corrections to b --\ensuremath{>} c transitions at zero recoil}},\ }\href
  {https://doi.org/10.1103/PhysRevLett.76.4124} {\bibfield  {journal} {\bibinfo
   {journal} {Phys. Rev. Lett.}\ }\textbf {\bibinfo {volume} {76}},\ \bibinfo
  {pages} {4124} (\bibinfo {year} {1996})},\ \Eprint
  {https://arxiv.org/abs/hep-ph/9603261} {arXiv:hep-ph/9603261} \BibitemShut
  {NoStop}%
\bibitem [{\citenamefont {Czarnecki}\ and\ \citenamefont
  {Melnikov}(1997)}]{Czarnecki:1997cf}%
  \BibitemOpen
  \bibfield  {author} {\bibinfo {author} {\bibfnamefont {A.}~\bibnamefont
  {Czarnecki}}\ and\ \bibinfo {author} {\bibfnamefont {K.}~\bibnamefont
  {Melnikov}},\ }\bibfield  {title} {\bibinfo {title} {{Two loop QCD
  corrections to b ---\ensuremath{>} c transitions at zero recoil: Analytical
  results}},\ }\href {https://doi.org/10.1016/S0550-3213(97)00500-2} {\bibfield
   {journal} {\bibinfo  {journal} {Nucl. Phys. B}\ }\textbf {\bibinfo {volume}
  {505}},\ \bibinfo {pages} {65} (\bibinfo {year} {1997})},\ \Eprint
  {https://arxiv.org/abs/hep-ph/9703277} {arXiv:hep-ph/9703277} \BibitemShut
  {NoStop}%
\bibitem [{\citenamefont {Franzkowski}\ and\ \citenamefont
  {Tausk}(1998)}]{Franzkowski:1997vg}%
  \BibitemOpen
  \bibfield  {author} {\bibinfo {author} {\bibfnamefont {J.}~\bibnamefont
  {Franzkowski}}\ and\ \bibinfo {author} {\bibfnamefont {J.~B.}\ \bibnamefont
  {Tausk}},\ }\bibfield  {title} {\bibinfo {title} {{O(alpha-s**2) corrections
  to b ---\ensuremath{>} c decay at zero recoil}},\ }\href
  {https://doi.org/10.1007/s100520050296} {\bibfield  {journal} {\bibinfo
  {journal} {Eur. Phys. J. C}\ }\textbf {\bibinfo {volume} {5}},\ \bibinfo
  {pages} {517} (\bibinfo {year} {1998})},\ \Eprint
  {https://arxiv.org/abs/hep-ph/9712205} {arXiv:hep-ph/9712205} \BibitemShut
  {NoStop}%
\bibitem [{\citenamefont {Bourrely}\ \emph {et~al.}(2009)\citenamefont
  {Bourrely}, \citenamefont {Caprini},\ and\ \citenamefont
  {Lellouch}}]{Bourrely:2008za}%
  \BibitemOpen
  \bibfield  {author} {\bibinfo {author} {\bibfnamefont {C.}~\bibnamefont
  {Bourrely}}, \bibinfo {author} {\bibfnamefont {I.}~\bibnamefont {Caprini}},\
  and\ \bibinfo {author} {\bibfnamefont {L.}~\bibnamefont {Lellouch}},\
  }\bibfield  {title} {\bibinfo {title} {{Model-independent description of B
  ---> pi l nu decays and a determination of |V(ub)|}},\ }\href
  {https://doi.org/10.1103/PhysRevD.82.099902, 10.1103/PhysRevD.79.013008}
  {\bibfield  {journal} {\bibinfo  {journal} {Phys. Rev.}\ }\textbf {\bibinfo
  {volume} {D79}},\ \bibinfo {pages} {013008} (\bibinfo {year} {2009})},\
  \bibinfo {note} {[Erratum: Phys. Rev. D82, 099902 (2010)]},\ \Eprint
  {https://arxiv.org/abs/0807.2722} {arXiv:0807.2722 [hep-ph]} \BibitemShut
  {NoStop}%
\bibitem [{\citenamefont {Falk}\ and\ \citenamefont
  {Neubert}(1993{\natexlab{b}})}]{Falk:1992wt}%
  \BibitemOpen
  \bibfield  {author} {\bibinfo {author} {\bibfnamefont {A.~F.}\ \bibnamefont
  {Falk}}\ and\ \bibinfo {author} {\bibfnamefont {M.}~\bibnamefont {Neubert}},\
  }\bibfield  {title} {\bibinfo {title} {{Second order power corrections in the
  heavy quark effective theory. 1. Formalism and meson form-factors}},\ }\href
  {https://doi.org/10.1103/PhysRevD.47.2965} {\bibfield  {journal} {\bibinfo
  {journal} {Phys. Rev. D}\ }\textbf {\bibinfo {volume} {47}},\ \bibinfo
  {pages} {2965} (\bibinfo {year} {1993}{\natexlab{b}})},\ \Eprint
  {https://arxiv.org/abs/hep-ph/9209268} {arXiv:hep-ph/9209268} \BibitemShut
  {NoStop}%
\bibitem [{\citenamefont {Neubert}(1993)}]{Neubert:1993iv}%
  \BibitemOpen
  \bibfield  {author} {\bibinfo {author} {\bibfnamefont {M.}~\bibnamefont
  {Neubert}},\ }\bibfield  {title} {\bibinfo {title} {{Reparametrization
  invariance and the expansion of currents in the heavy quark effective
  theory}},\ }\href {https://doi.org/10.1016/0370-2693(93)90091-U} {\bibfield
  {journal} {\bibinfo  {journal} {Phys. Lett. B}\ }\textbf {\bibinfo {volume}
  {306}},\ \bibinfo {pages} {357} (\bibinfo {year} {1993})},\ \Eprint
  {https://arxiv.org/abs/hep-ph/9302269} {arXiv:hep-ph/9302269} \BibitemShut
  {NoStop}%
\bibitem [{\citenamefont {Neubert}(1994)}]{Neubert:1993mb}%
  \BibitemOpen
  \bibfield  {author} {\bibinfo {author} {\bibfnamefont {M.}~\bibnamefont
  {Neubert}},\ }\bibfield  {title} {\bibinfo {title} {{Heavy quark symmetry}},\
  }\href {https://doi.org/10.1016/0370-1573(94)90091-4} {\bibfield  {journal}
  {\bibinfo  {journal} {Phys. Rept.}\ }\textbf {\bibinfo {volume} {245}},\
  \bibinfo {pages} {259} (\bibinfo {year} {1994})},\ \Eprint
  {https://arxiv.org/abs/hep-ph/9306320} {arXiv:hep-ph/9306320} \BibitemShut
  {NoStop}%
\end{thebibliography}
\end{document}